\documentclass{elsarticle}

\pdfoutput=1
\usepackage{lineno,hyperref}

\usepackage{amsmath}
\usepackage{booktabs}

\usepackage{multirow}

\usepackage{booktabs}
\usepackage{caption} 
\modulolinenumbers[5]

\journal{Computers and fluids}

\usepackage{numcompress}\biboptions{authoryear}

\begin{document}

\begin{frontmatter}

\title{A high-resolution DNS study of compressible flow past an LPT blade in a cascade}

\author{Rajesh Ranjan, S M Deshpande} 
\author{Roddam Narasimha\corref{mycorrespondingauthor}}

\address{JNCASR, Bangalore, India}


\cortext[mycorrespondingauthor]{Corresponding author}
\ead{roddam@jncasr.ac.in}


\begin{abstract}
Flow past a low pressure turbine blade in a cascade at $Re \approx 52000$ and angle of incidence $\alpha = 45.5^{0}$ is solved using a code developed in-house for solving 3D compressible Navier-Stokes equations. This code, named ANUROOP, has been developed in the finite volume framework using kinetic energy preserving second order central differencing scheme for calculating fluxes, and is compatible with  hybrid grids. ANUROOP was verified and validated against several test cases with Mach numbers ranging from 0.1 (Taylor-Green vortex) to 1.5 (compressible turbulent channel flow). The code was found to be robust and stable, and the kinetic energy decay obeys the compressible Navier-Stokes equations.

A hybrid grid, with a high resolution hexahedral orthogonal mesh in the boundary layer and unstructured (also hexahedral) elements in the rest of the domain is used for the turbine blade simulation.  Total grid size (160 million) is approximately an order of magnitude higher than in previous simulations for the same flow conditions and using similar numerical methods. The discrepancy in the pressure distribution in earlier studies compared to experimental data has been removed in this simulation. The trailing edge separation bubble has been characterized and a detailed discussion on the effect of surface curvature is presented.     
\end{abstract}

\begin{keyword}
DNS,
LPT blade,
Separation bubble,
Curvature effects
\end{keyword}

\end{frontmatter}


\section{Introduction}
For realistic, complex engineering flows, direct numerical simulation (DNS) is yet to become an established analysis tool because of computational domains with complicated geometries, and the very high computational power and memory requirement when Reynolds number ($Re$) is  high ($\mathcal{O}(10^6)$ based on a typical length scale (chord etc.) and free-stream velocity). In the engineering literature, the flow past low pressure (LP) gas turbine blades in a cascade is perhaps one of the practical engineering problems where DNS has been most widely explored \citep{wu2001, michelassi2002, wissink2003, kalitzin2003dns, wissink2006, wissink2006direct, wissink_jfm2006, rajesh2013numerical, ranjan2014direct, rajesh2015IUTAM, michelassi2015compressible, garai2015dns}. 
This is because of the relatively lower computational power requirement Reynolds numbers ($Re \approx 10^4-10^5$ in some of the LPT flows),  but  by a degree of flow complexity that puts it beyond the RANS codes in common use. A detailed survey of the computational studies on LPT blades is presented in \cite{rajeshthesis2015} and briefly described in later sections.

Among these studies, most of the simulations (except the recent ones \cite{rajesh2013numerical, michelassi2015compressible, garai2015dns} ) solve incompressible Navier-Stokes equations (thereby ignoring compressibility effects), and are performed using the Finite Volume Method (FVM) which has the advantage of being able to work on grids in computational domains with complicated geometry. The primary focus of these studies is to understand the effect of free-stream turbulence, wake  from upstream rotor stages etc. on the general features of the flow, but there has been no detailed study of the blade boundary layer. This may be due to the elliptic grid approach used in these studies and the associated restrictions in adequate representation and resolution of the boundary layer. Furthermore, most of these studies were performed when sufficiently powerful computers were not available.   

In this paper, we describe a new code that has been developed to exploit high performance computing (HPC) to do a detailed study of the boundary layer on an LPT blade in a cascade. This code, named ANUROOP, has been thus developed at JNCASR  to solve the 3D compressible Navier-Stokes equations in 3D space and time. The decision to solve the blade flow using compressible formulations was influenced by the fact that some of the earlier studies reported above have suspected compressibility effects as one of the reasons for the discrepancy between experimental and computational results. Further when the development of ANUROOP was started in 2010, no compressible DNS past LP turbine blades had been reported in the open literature to the best of the authors' knowledge. Though ANUROOP is developed primarily to simulate flow past gas turbine blades, the general treatment of the grid, boundary conditions and flux calculations make this DNS code a general computational tool to study a class of realistic engineering flows.     
 
This paper is chiefly divided into two parts. In the first part, the methodology of ANUROOP is described in detail along with a brief description of test cases used for the validation of the code. In the second part, study on a high lift low pressure turbine blade T106A using this code is presented. The present grid strategy in this simulation, which resolves the boundary layer mesh upto sufficient resolution without unduly increasing the grid size,  is described. More emphasis is given on the streamline resolution and hence the curvature of the blade in order to understand their effect on the boundary layer parameters.    

\part{Code ANUROOP}
\section{Governing Equations and Methodology}
The Navier-Stokes equations are the governing equations in fluid flow in aircraft gas turbines and express conservation of  mass, momentum, and energy. ANUROOP solves these equations in compressible form as described below:

\begin{subequations}
\label{eqn:govern}
\begin{align}
\frac{\partial{\rho}}{{\partial{t}}}   +  \frac{\partial{(\rho u_{j})}}{\partial{x_{j}}} &= 0 \\
\frac{\partial{(\rho u_{i})}}{\partial{t}}+\frac{\partial{(\rho
u_{i} u_{j})}}{\partial{x_{i}}} &= -
\frac{\partial{p}}{\partial{x_{i}}} +
\frac{\partial{\tau_{ij}}}{\partial{x_{j}}} \\
\frac{\partial{(\rho E)}}{\partial{t}}+\frac{\partial{(\rho u_{j}
H)}}{\partial{x_{j}}} &= 
\frac{\partial}{\partial{x_{j}}} {(u_{i} \tau_{ij})}
-\frac{\partial} {\partial{x_{j}}} q_{j}
\end{align}
\end{subequations}
where $\rho$ and $u_{i}$ are density and velocity vector components ($i ~= ~1,2,3$) respectively; $E$ and $H$ are total energy and total enthalpy per unit mass respectively, given by  
\begin{align*}
\displaystyle
E &= e + K = c_{v}T + \sum_{i=1}^{3} \frac{u_{i}u_{i}}{2} \\
H &= E + \frac{p}{\rho} 
\end{align*}
where $e$ is the internal energy and $K$ is the kinetic energy of the gas. 
To close the above set of equations, they must be supplemented by an equation of state which, for a perfect gas, is:
\begin{align}
\displaystyle
p &= \rho R  T 
\end{align}
The viscous stress tensor and the heat flux vector are given respectively by 
\begin{align}
\displaystyle \tau_{ij} &= \mu\left [ \frac{\partial{u_{i}}}{\partial{x_ {j}}} +
\frac{\partial{u_{j}}}{\partial{x_ {i}}} - \frac{2}{3} \delta_{ij}
\frac{\partial{u_{k}}}{\partial{x_{k}}} \right ] \\
\displaystyle q_{i} &= -k \frac{dT}{d{x_{i}}}
\end{align}
The co-efficients $\mu$ and $\kappa$ are viscosity and thermal conductivity respectively, which vary with the local temperature $T$ and are assumed to follow Sutherland's law \citep{sutherland1893lii}.

The above equations are solved in non-dimensional form in ANUROOP, in which the quantities are non-dimensionalized using appropriate length and velocity scales. For example, in the turbine blade simulation,  the axial chord length and free-stream velocity were taken as length and velocity scales respectively.

\section{The Flow-solver Algorithm}
In ANUROOP, the underlying governing equations (1) to (4) are solved using cell-centered finite volume method (FVM). FVM is flexible, robust and allows the solution of flow problems in domains with a complicated geometry.  The flux calculations in FVM are local and no separate formulation is needed for structured and unstructured grids. This is useful for solving wall-bounded flows past a complicated surface geometry where different grid topologies can be used for the boundary layer and the external flow. Further, since FVM solves the equations in conservative form, it is useful for solving compressible flow equations where conservation of energy is a prime requirement.

\subsection{Spatial Discretization}
The computation of inviscid and viscous fluxes at grid points requires numerical schemes which are not only sufficiently accurate but also stable and robust.  Viscous terms generally add stability to the equations and hence do not require any special treatment, except for a suitable method to calculate gradients in an unstructured grid setup. Calculation of inviscid terms however involves in most methods solving the Riemann problem and hence requires special treatment.  

\subsubsection*{Inviscid Flux Calculation}
For high fidelity simulations like DNS, we require numerical schemes that are non-dissipative and have small aliasing errors, and yet stable and robust. Upwind schemes that use biased-differencing, based on the directions of characteristic waves, are stable and robust, but generally dissipative, and hence not considered a suitable choice for DNS. Central-difference schemes, on the other hand, are non-dissipative but have been found to be generally unstable because they do not conserve kinetic energy in the discrete sense. 

ANUROOP uses a variant of the central-differencing scheme, known as kinetic energy preserving (KEP) scheme, as described in  \cite{jameson2008}. In this scheme, the kinetic energy is discretely preserved in the implementaton for compressible Navier-Stokes equations and is found to be robust in various 2D DNS studies, including the shock tube \citep{jameson2008, allaneau2009direct}, plunging airfoils \citep{allaneau2010direct}, and flow past circular cylinder \citep{shoeybi2010}. In ANUROOP, the KEP scheme is implemented for solving three-dimensional flow past a turbine blade on a hybrid grid. 

\subsubsection*{Viscous Flux Calculation}
\begin{figure}
\centering
 \includegraphics[trim=0 0 0 0, clip, width=0.4\linewidth]{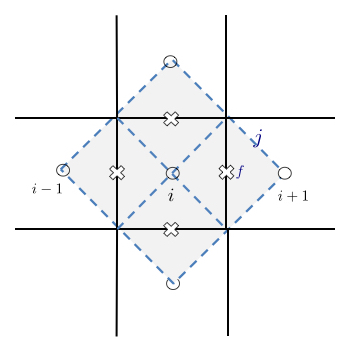}
 \caption{Co-volume (dashed lines) for calculation of gradients at face-centers}
\label{fig:covolume}
\end{figure}

In the cell-centered finite-volume scheme, the calculation of the viscous and conduction terms requires gradients of velocity and temperature at the face of the volume element.  In ANUROOP,  these face gradients are calculated using the Green-Gauss theorem. An auxilliary volume (also called co-volume) is formed around each face connecting the cell-centroids and the face nodes as shown in Fig. ~\ref{fig:covolume}. At the boundaries half co-volumes are constructed; faces of this co-volume are called co-faces. Now using the Green-Gauss theorem, the gradient of a quantity $\Phi$ at a face $f$ will be given by
\begin{equation}
 {\mathbf{\nabla}} \Phi_{f} = \frac{\Sigma \Phi_{j} {\mathbf{A_{j}}}}{\Omega_{f}}  
\end{equation}
where $\Omega_{f}$ is the volume of the co-volume encircling face $f$; $\Phi_{j}$ and $\mathbf{A_{j}}$ are the averaged values of $\Phi$ at co-faces and the face-normal areas of the co-faces of the co-volume respectively.

To implement this scheme, values at the nodes (cell-vertices in this case) need to be calculated at every time-step as the physical variables are stored at the cell-centers in our approach. Typically, area-weighted average or inverse area-weighted average interpolations are used to get the values at the nodes. A more accurate method for interpolation on unstructured grids is the pseudo-Laplacian weighted average (see ~\cite{holmes1989solution} for details) as suggested by \cite{rausch1992spatial} and \cite{frink1994recent}. In ANUROOP, the pseudo-Laplacian weighted average is used for the cell to node interpolation except at the boundaries where the area-weighted average replaces it. At the boundaries there is not enough connectivity information to accurately calculate the pseudo-Laplacian weights for averaging.

\subsection{Time Discretization}
\cite{subbareddy2009} have proposed a fully discrete kinetic energy consistent finite-volume scheme, where an implicit extension of the Crank-Nicholson scheme has been used to conserve kinetic energy in time. However, this proposal entails substantial increase in the cost of computing fluxes as well as higher storage requirements.

A more economical option is to use a total-variation-diminishing (TVD) scheme that does not allow kinetic energy to grow rapidly. Runge-Kutta 3 (RK-3) scheme based on \cite{shu1988efficient} is one such scheme that has been used widely because of lower computation and storage requirements. In ANUROOP, this scheme is implemented as it is easily parallelizable and also more scalable \citep{allaneau2009direct}, and hence is suitable for DNS studies.  The implementation of RK-3 is as follows: 
For the equation 
\begin{eqnarray*}
\displaystyle
\frac{dU}{dt}+R(U)&=& 0
\end{eqnarray*}
RK-3 takes
\begin{eqnarray*}
\displaystyle
U^{(1)} &=& U^{(n)} - \Delta t ~ R(U^{(n)})\\
U^{(2)} &=& \frac{3}{4} U^{(n)} + \frac{1}{4} U^{(1)} - \frac{1}{4} \Delta t ~R(U^{(1)})\\
U^{(n+1)} &=& \frac{1}{3} U^{(n)} + \frac{2}{3} U^{(1)} - \frac{2}{3} \Delta t ~R(U^{(2)})
\end{eqnarray*}

\subsubsection*{Calculation of Time Step}
In ANUROOP, the maximum allowable time step is calculated using the Courant-Friedrichs-Lewy (CFL) condition. This condition ensures that the flow on the stencil of a grid respects the physics. For the 1D Euler equations, the CFL condition for time step is given by
\begin{eqnarray*}
\displaystyle
\Delta t  &=& \sigma ~ \frac{\Delta x}{\lvert \Lambda \rvert}
\end{eqnarray*}
where $\displaystyle \Delta x/{\lvert \Lambda \rvert}$ is the time needed for information to propagate on a grid of size $\Delta x$ with velocity $\Lambda$, and $\sigma$ is a positive coefficient, known as the CFL number. For Euler flows, $\Lambda$ corresponds to the maximum eigen value in the convective flux Jacobian. For viscous flows, the spectral radius of the viscous flux Jacobian needs to be included in the calculation of maximum $\Delta t $ as the flow in the boundary layer can severely restrict the maximum allowable time-step.

For 3D viscous calculations in ANUROOP on unstructured grids,  the time-step computation follows the method suggested by \cite{blazek2005computational},
\begin{eqnarray}
\displaystyle
\Delta t_i  &=& \sigma ~ \frac{\Omega_i}{(\Lambda_{inv} + C ~ \Lambda_v)_i}
\end{eqnarray}
Here $\Lambda_{inv}$ and $\Lambda_{v}$ represent a sum of Euler and viscous spectral radii of all the faces over all control volumes, and are given by:
\begin{eqnarray*}
\displaystyle
(\Lambda_{inv})_i &=& \sum_{J=1}^{N_F} (\lvert \mathbf{v} \cdot \mathbf{n} \rvert  +  c ) ~  \Delta S_{J}\\
(\Lambda_{v})_i &=& \frac{1}{\Omega_i} \sum_{J=1}^{N_F}  \bigg[\mathrm{max} \bigg(\frac{4}{3\rho}, \frac{\gamma}{\rho} \bigg) 
\bigg(\frac{\mu}{Pr} \bigg) (\Delta S_{J})^2 \bigg]
\end{eqnarray*}
Here $\mathbf{v} \cdot \mathbf{n}$ and $c$ are the normal velocity and speed of sound on face $J$ of cell $i$ respectively; $\Delta S_{J}$ and $\Omega_i$ are geometrical parameters representing area of face $J$ and volume of cell $i$ containing all such faces. 

The constant $C$ that multiplies the viscous spectral radius $ \Lambda_v$ is taken as 4 (recommended for central schemes in \cite{blazek2005computational}), and the CFL number is kept below 1  for the time-stepping through out the computation. 

\section{Boundary Conditions}
Inlet, outlet, wall and periodic boundary conditions, which are needed to simulate the flow between consecutive blades, are implemented in ANUROOP. The values of primitive variables at the boundary are updated every time-step using the boundary conditions.  All the boundary conditions are imposed by adding ghost cells. For periodic boundary conditions, the values at the first interior cells of the first boundary are copied to corresponding ghost cells in the other periodic boundary and vice versa.  

For wall, inlet and outlet,  the details of the boundary conditions implemented in ANUROOP are given below.
\subsection{Wall Boundary Condition}
For the wall a mirror boundary condition has been used, where the variables in the interior cells next to the boundary are mirrored appropriately to corresponding ghost cells and an average is taken to get the value at the wall face. For a no-slip stationary isothermal wall, the ghost values are as follows:
\begin{eqnarray*}
\displaystyle
u_g &=& - u_i \\
v_g &=& - v_i \\
w_g &=& - w_i \\
T_g &=& 2\times T_w - T_i
\end{eqnarray*}
where subscript $i$ and $g$ stand for interior and ghost cells respectively. $T_w$ is the imposed wall temperature. The density is obtained using the continuity equation.

\subsection{Inlet/Outlet Boundary Conditions}
The treatment of flow at the inlet and outlet is critical, since in the simulation an infinite flow domain is restricted to a finite computational domain. An improper boundary condition may lead to unphysical oscillations near the boundary, which with progress in time may either lead to blow up of the code or give spurious results. It is necessary to avoid the reflection of the outgoing waves (from the computational domain) to ensure smooth inflow/outflow. To do this, primitive variables at the boundary are extrapolated using characteristics variables (known as Riemann invariants), as given in ~\cite{hirsch1988numerical}. 

The Riemann invariants are calculated based on the values at the interior cells of the boundaries or the far-field values, depending on whether the characteristics point inwards or outwards.

For 3D Euler equations there are 5 eigen values (characteristics) of the Jacobian matrix, given by
\begin{subequations}
\begin{align}
\displaystyle \lambda_1 ~ &=  u_{\perp}+c \\
\displaystyle \lambda_{2,3,4} ~ &=  u_{\perp} \\
\displaystyle \lambda_5 ~ &=  u_{\perp}-c
\end{align}
\end{subequations}
where the subscript $\perp$ indicates the velocity normal to the boundary face, given by $\displaystyle \mathbf{v} \cdot \mathbf{n}$.

These eigen values determine whether the wave is entering or leaving the domain. At the subsonic inlet  ($-1 \le M_{\perp} < 0; ~ u_{\perp} < 0, c > 0, u_{\perp} < c $), there is one right running wave ($u_{\perp}+c > 0$) and two left-running waves ($u_{\perp} < 0, ~ u_{\perp} - c < 0$).  At the subsonic outlet  ($0 \ge M_{\perp} < 1; ~ u_{\perp} > 0, ~ c > 0, ~ u_{\perp} < c $), there are two right running waves ($u_{\perp} > 0, ~ u_{\perp}+c > 0$) and one left-running wave ($u_{\perp} - c < 0$). The invariants corresponding to the right running wave are calculated based on the values at the interior cells and those corresponding to the left-running wave are calculated based on far-field values, as given below.\\

\textbf{Subsonic inflow boundary}
\begin{subequations}
\begin{align}
\displaystyle \psi_1 ~ &\equiv ~ \psi_{1i}  ~= ~ \frac{2c_i}{\gamma -1} + u_{\perp i}\\
\psi_2 ~ &\equiv ~ \psi_{2\infty} = ~ \frac{p_{\infty}}{\rho_{\infty}^{\gamma}} \\
\psi_3  ~ &\equiv ~ \psi_{3\infty} = ~ \frac{2c_{\infty}}{\gamma -1} - u_{\perp \infty} 
\end{align}
\end{subequations}

\textbf{Subsonic outflow boundary}
\begin{subequations}
\begin{align}
\displaystyle \psi_1 ~ &\equiv ~ \psi_{1i}  ~= ~ \frac{2c_i}{\gamma -1} + u_{\perp i}\\
\psi_2 ~ &\equiv ~ \psi_{2i} ~= ~ \frac{p_{i}}{\rho_{i}^{\gamma}} \\
\psi_3  ~ &\equiv ~ \psi_{3\infty} = ~ \frac{2c_{\infty}}{\gamma -1} - u_{\perp \infty}
\end{align}
\end{subequations}
After the invariants are known, the values at the boundary face are thus extrapolated as
\begin{align*}
\displaystyle u_{\perp bf} ~=& ~\frac{\psi_1 + \psi_3}{2}\\
\displaystyle c_{bf} ~=& ~\bigg( \frac{\gamma -1}{4} \bigg) (\psi_1 - \psi_3)\\
\displaystyle \rho_{bf}~ =& ~\bigg(\frac{a_f^2}{\gamma \psi_2}\bigg)^{\frac{1}{\gamma - 1}}\\
\displaystyle p_{bf} ~=&~ \frac{\rho_f a_f^2}{\gamma} 
\end{align*}
$u_{\perp bf}$ is then again back-transformed to get all 3 components of velocity at the boundary. 

For the simulations with inflow turbulence, the turbulent fluctuations are superimposed on the mean value of the velocities as obtained from the boundary condition. 

\section{Validation Studies}
Systematic validation studies were performed to test the ability of ANUROOP code for 1D, 2D and 3D flows (\cite{rajeshthesis2015}). We present here only two 3D studies: the Taylor-Green vortex and supersonic turbulent channel flow, where the code is tested for accuracy, stability and robustness.  

\subsection{The Taylor-Green vortex}
The Taylor-Green vortex is the unsteady flow problem of a decaying vortex in a box proposed by G. I. Taylor and George Green.  The flow inside a periodic box is initialized by simple perturbations of sines and cosines with zero-mean, representing counter-rotating vortices, which decay with time at a rate governed by viscosity. An exact closed form solution can be constructed in 2D in the incompressible limit.
\begin{figure}
\centering
\includegraphics[trim=0 0 0 0, clip, width=0.7\linewidth, angle=0]{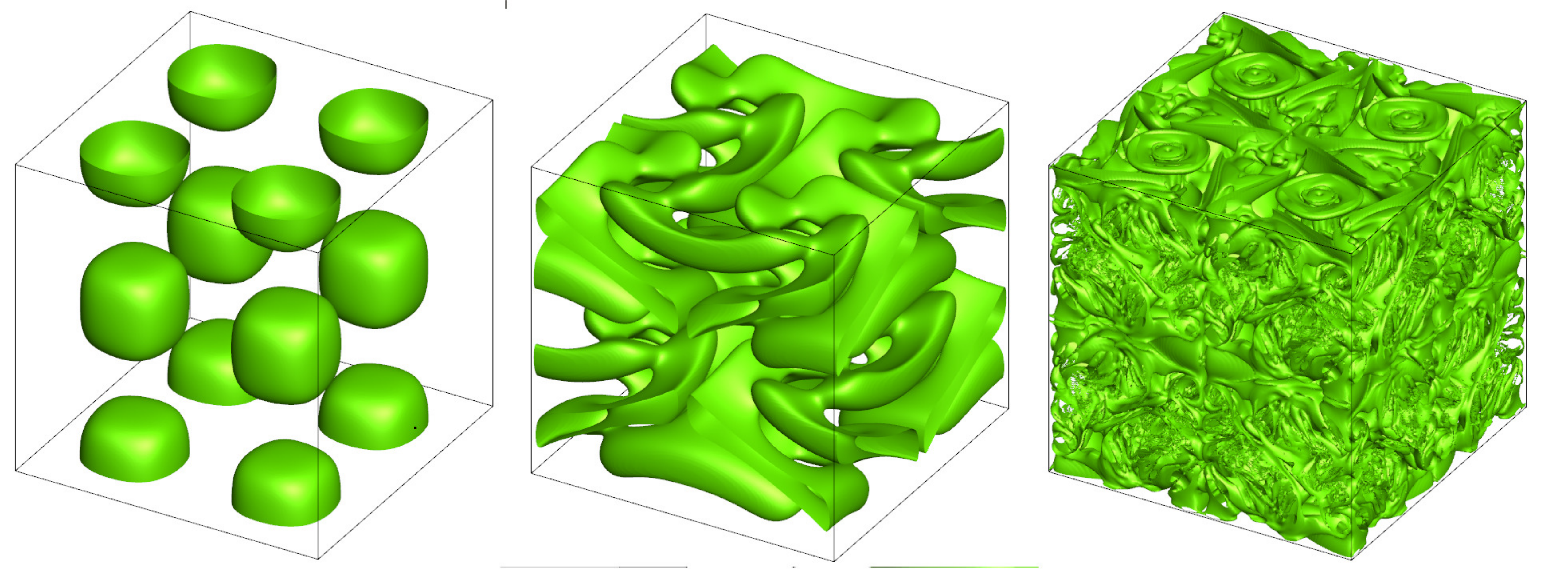}
\caption{Iso-surfaces of vorticity magnitude at $t$ = 0.5, 2.0 and 13.0}
\label{fig:TG1}
\end{figure}
The same problem in 2D can be extended to 3-D flows in the weakly compressible limit.  Figure \ref{fig:TG1} shows the stages of the flow as time progresses in the present simulation.  The pressure is initialized, as above, at a value corresponding to the solution of the pressure Poisson equation. The flow first goes transitional and later becomes fully turbulent, with the generation of small scales. Since there is no energy input, the flow soon starts decaying as in unforced homogeneous turbulence. 

This is a good validation problem for a DNS code with simple initial and boundary conditions. In the present study, 3D DNS was performed to assess the robustness and temporal accuracy of the ANUROOP code, particularly the ability of the KEP scheme to capture the evolution of kinetic energy accurately. 

\subsubsection*{Computational Setup}
The Taylor-Green vortex  flow is the evolution of a rotational velocity field in a triply-periodic cube $0 \le x,y,z \le 2\pi$, from the initial conditions:
\begin{eqnarray*}
u(0; x,y,z) &=& U_{\textrm{ref}}\ \mathrm{sin}x \ \mathrm{cos}y \ \mathrm{cos}z\\
v(0; x,y,z) &=& -U_{\textrm{ref}}\ \mathrm{cos}x \ \mathrm{sin}y \ \mathrm{cos}z\\
w(0; x,y,z) &=& 0\\
\rho(0; x,y,z) &=& \rho_{\textrm{ref}}\\
p(0; x,y,z) &=& p_{0} + \frac{\rho_{\textrm{ref}}U_{\textrm{ref}}^2}{16}(\mathrm{cos}2x + \mathrm{cos}2y)(\mathrm{cos}2z + 2)
\end{eqnarray*}
Flow parameters set for the simulation are $Re = 1600$ based on the box length and the speed of sound, and $M = 0.1$. The initial Mach number $M$ gives the value of $p_{0}$. The grid chosen for the current simulation is $256^3$. The flow has been allowed to develop till $t = 13.0$, where $t$ is the non-dimensional time based on $\displaystyle L_{ref}/U_{ref}$. This set-up allows comparisons with the earlier benchmark DNS \citep{hiocfd2013} performed using  a psuedo-spectral method with $512^3$ grid.

\subsubsection*{Results and Discussion}   
\def\figsubcap#1{\par\noindent\centering\footnotesize(#1)}
\begin{figure}
  \hspace*{4pt}
      \parbox{2.3in}{\includegraphics[scale=0.14]{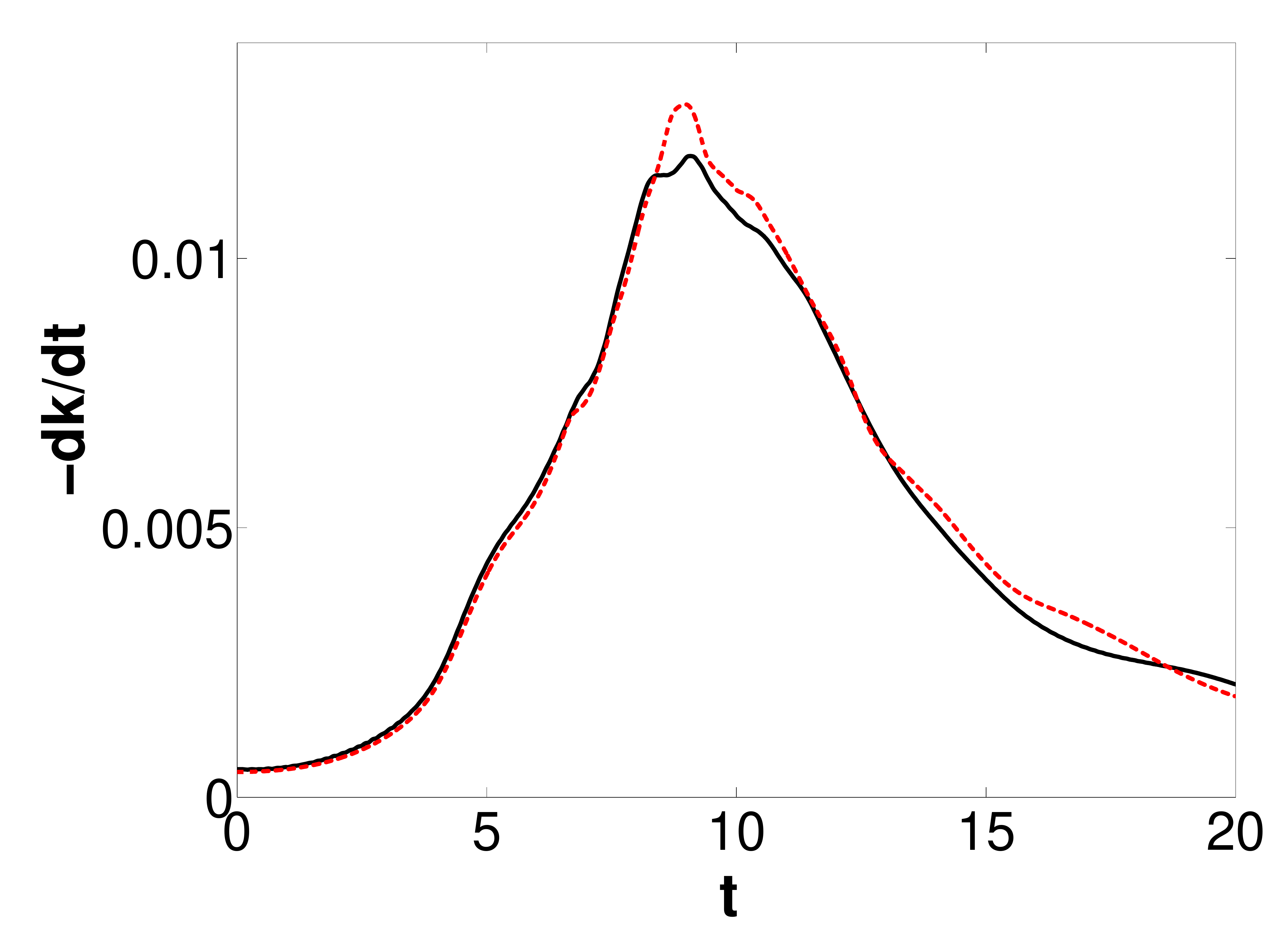}\figsubcap{a}}
\hspace*{4pt}
      \parbox{2.3in}{\fbox{\includegraphics[scale=0.16]{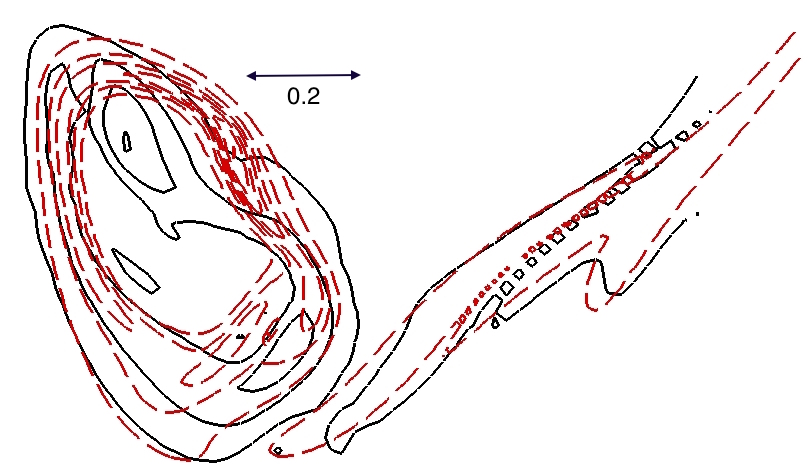}}\figsubcap{b}}
\caption{(a) Kinetic energy dissipation rate (b) Contours of non-dimensional vorticity magnitude of 5, 10 and 15 at $t = 8.0$ on one of the periodic faces ($x = 0$). Results are compared with a $512^3$ pseudo-spectral simulation \cite{hiocfd2013}}
\label{fig:TG3}
\end{figure}

Figures ~\ref{fig:TG1} and ~\ref{fig:TG3} present results for the $Re=1600$ simulation. Figure ~\ref{fig:TG1} shows iso-surfaces of vorticity magnitude ($|\bar{\omega}|$) as the flow evolves in time. It illustrates the cascade process, in which multiple scales are generated from a single initial large scale into very small scales due to non-linear interaction of eddies. 

The kinetic energy of the system decays due to viscous dissipation.  Figure ~\ref{fig:TG3}(a) shows the rate of decay of kinetic energy as compared with the benchmark pseudo-spectral DNS results by \cite{hiocfd2013} on a $512^3$ grid. The overall match is good and the maximum decay rate at $t \approx 9$ is captured. The peak on the decay rate predicted in the current simulation is slightly lower compared to the reference simulation. This may be due to the fact that the grid size used in current simulation is one-eighth of that used in the pseudo-spectral simulation. The temporal resolution here is also governed by the grid resolution. Figure ~\ref{fig:TG3}(b) shows contours of the non-dimensional vorticity magnitude on one of the periodic faces at $t\approx9.0$  when dissipation is near maximum. The structures obtained in the current simulation are strikingly similar to those obtained by \cite{hiocfd2013}.

Several other test simulations have been made over a range of $Re = 1000-3000$ to check the stability of the present semi kinetic-energy preserving central scheme. The simulations progressed without any difficulty in all cases, confirming the robustness of the code.    
\subsection{Supersonic Turbulent Channel Flow}
DNS of compressible channnel flow has been performed to test the ANUROOP code against wall-bounded flows at finite Mach number. Past studies by \cite{coleman1995}  for an isothermal supersonic channel flow offer a good case for validation of the present code. Two plates separated by a  width of $2H$ constitute the channel in which the fluid flows. The flow is statistically homogeneous in the streamwise ($x$) as well  as spanwise($z$)  directions, and is driven by a body-force that keeps the mass-flux constant. 

\subsubsection*{Problem Formulation}
The main parameters governing the flow are Mach number and Reynolds number. Of the various cases described in \cite{coleman1995}, the one chosen here for the validation is at $M=1.5$ and $Re=3000$ (based on centerline velocity, viscosity at wall temperature, and channel half-height $H$). Prandtl number $Pr$ is kept constant at 0.7 throughout the flow  and the viscosity of the fluid varies according to the power law $\mu \propto ({T/T_{w}})^{0.7}$, where $T_{w}$ is the prescribed (constant) wall temperature.

Body-force terms have been added as sources in the momentum and energy equations in Eqn. \ref{eqn:govern}. The body force is varied with time to keep stream-wise mass-flux constant. Thus, the change in body-force after every iteration is given by  

\begin{equation*}
  \Delta B = \frac{(\int \rho u dy dz)_{n} - (\int \rho u dy dz)_{n-1}} {(\int \rho dy dz)_{n}}
\end{equation*}

\subsubsection*{Geometry and Grid}
The sides of the computational domain are $\displaystyle L_{x} = 2\pi, L_{y} = 2, L_{z} = 4\pi/3$. There are 100 grid points chosen in each direction to make the grid. The grid is uniform in streamwise ($x$) and spanwise ($z$) directions but stretching is used along wall-normal direction $y$ to obtain a sufficiently fine grid to capture the near-wall flow. The stretching is done using the hyperbolic tangent function, as follows: 
\begin{equation*}
 y_{j} = \frac{\mathrm{tanh}[c(2(j-1)/(n_{y}-1)-1]}{\mathrm{tanh}(c)}
\end{equation*}
where $n_{y}$ is the total number of grid-points in the $y$-direction and $c$ is a constant taken as 1.7 (as in \cite{subbareddy2009}). The range of $y$ spans $-1$ to $+1$ with $y=0$ at the centerline of the channel. 

The simulation has been initialized with a parabolic laminar velocity profile  superimposed with white noise (with zero mean) perturbations. 

\subsubsection*{Results and Discussion}
Instantaneous quantities have been time- and span-averaged to get the mean values. Favre-averaging that accounts for density variations is used to obtain the Favre mean, defined as follows:
\[\tilde{\phi} = \frac{\langle \rho \phi \rangle}{\langle \rho \rangle} \]
Here the tilde and angular brackets denote Favre and Reynolds averaging respectively. 
The fluctations around the mean are given by:
\[ \phi^{''} = \phi - \tilde{\phi} \]

Mean streamwise velocity is shown in Fig. \ref{fig:channel_mean}, compared with the spectral simulation results of \cite{coleman1995}. The figure shows a good agreement between the two. 

\def\figsubcap#1{\par\noindent\centering\footnotesize(#1)}
\begin{figure}
  \hspace*{4pt}
    \parbox{2.5in}{\includegraphics[scale=0.25]{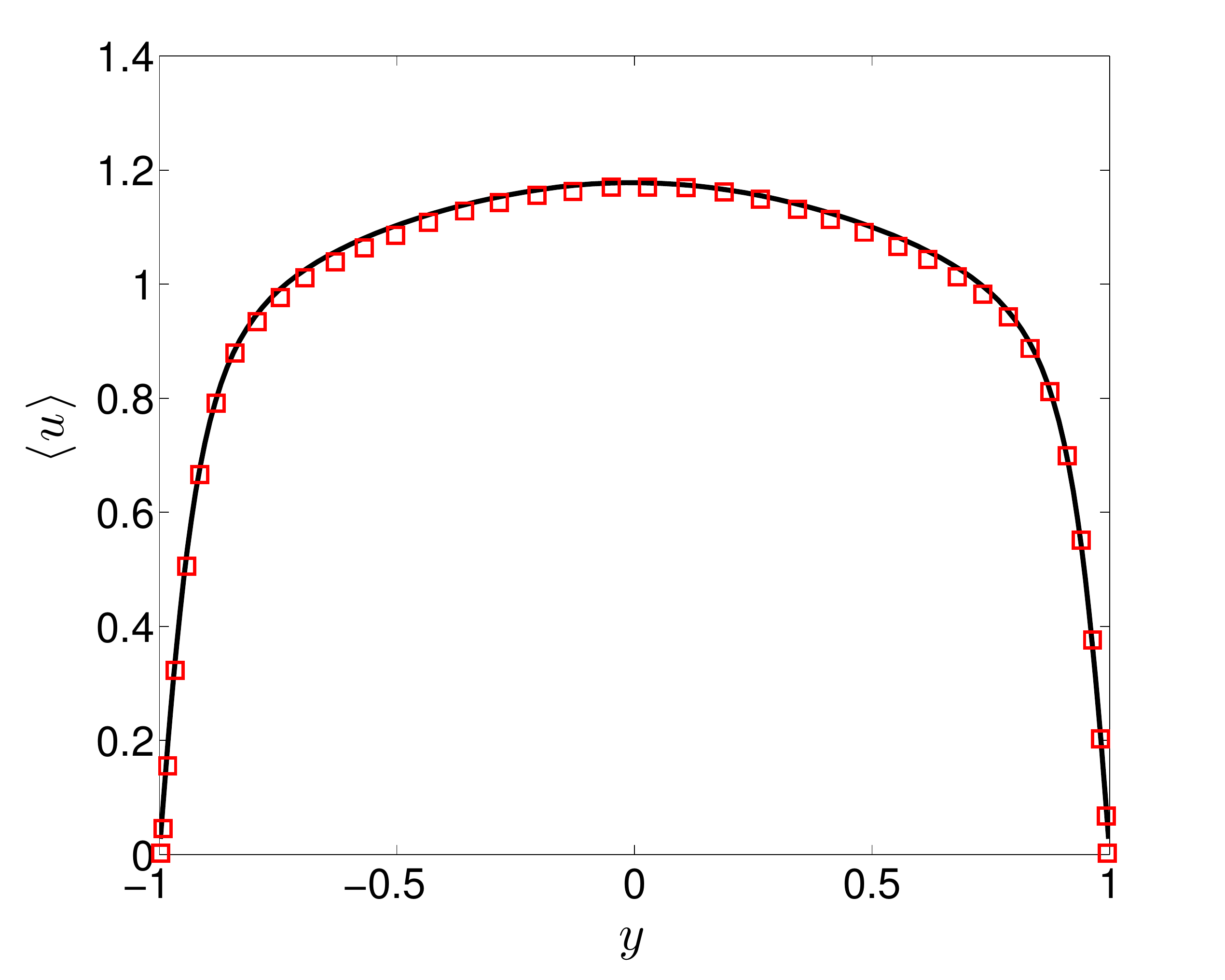}\figsubcap{a}}
  \hspace*{4pt}
    \parbox{2.3in}{\includegraphics[scale=0.22]{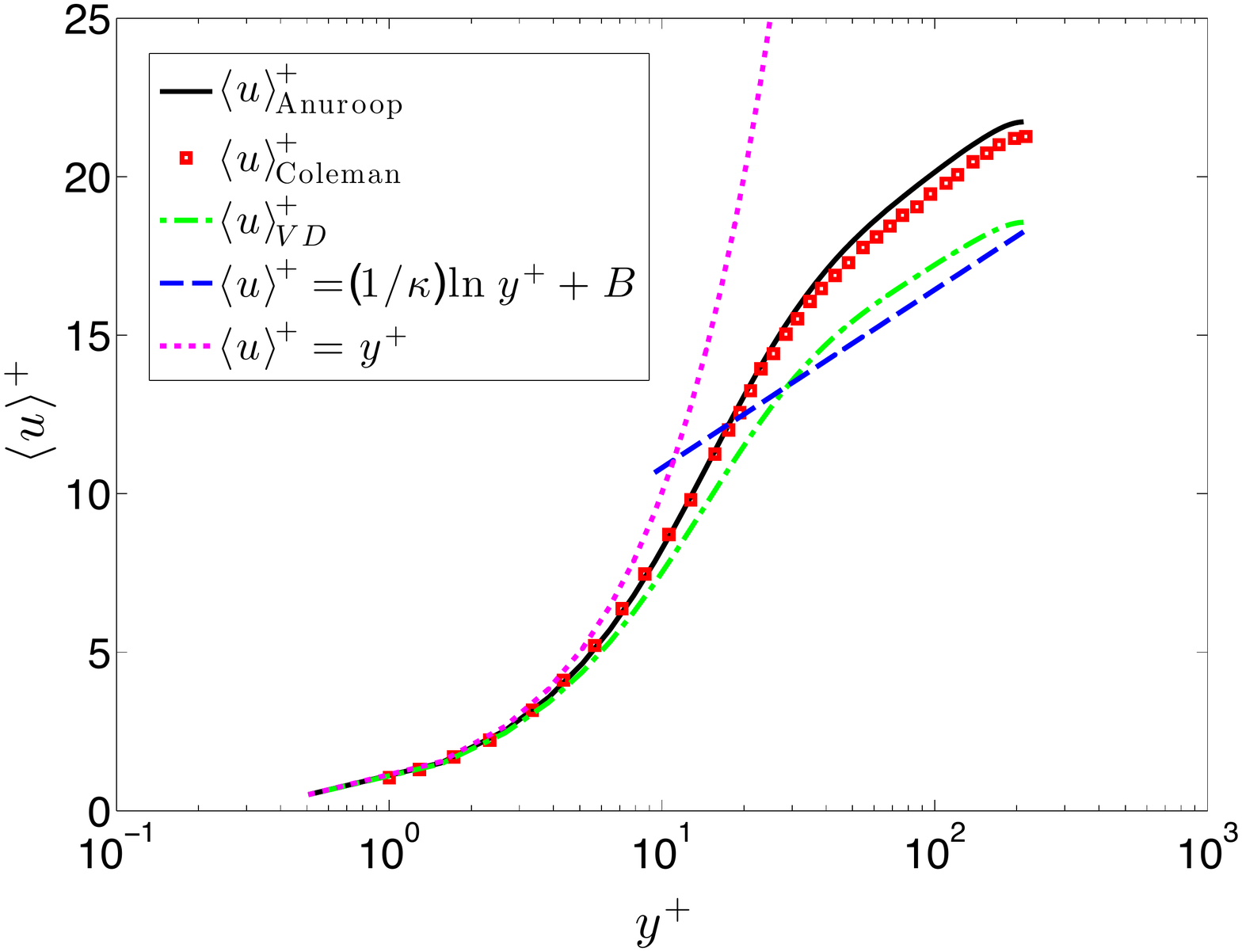}\figsubcap{b}}
  \caption{(a) Mean velocity profile. Symbols are values from \cite{coleman1995} simulation. (b) Mean streamwise velocity profile in wall variables and the Van-Driest transformation compared with incompressible law of the wall: $u_{wall} = y^+, u_{log} = \frac{1}{0.41}\mathrm{ln}y^+ + 5.2 $
}
\label{fig:channel_mean}
\end{figure}

Table ~\ref{channel_comparison} compares various parameters computed from mean results, against the \cite{coleman1995} simulation. The mean centerline velocity is 1.178 which is very close to the value 1.175 obtained by \cite{coleman1995}. Other variables listed in the table at the centerline as well as at the wall also show good match.
 
\begin{table}[t!]
\caption{Comparison of mean quantities with \cite{coleman1995} simulation}
\begin{center}
\label{channel_comparison}
\begin{tabular}{|c|c|c|c|c|c|c|c|c|}
\toprule
 \multirow{2}{*}{ \textbf{Case}} & \multicolumn{5}{|c|}{Centerline} & \multicolumn{3}{|c|}{Wall} \\[0.5em]
 \cline{2-9}
& $\langle u_{c} \rangle$ &  $\langle \rho_{c} \rangle$ & $\langle T_{c} \rangle$ &  $M_{c}$ &   $Re_{c}$ &  $\langle \rho_{w} \rangle$ & $M_{\tau}$ &   $Re_{\tau}$ \\[0.5em]
 \midrule
 ANUROOP & 1.178 & 0.979 & 1.391 & 1.498 & 2746 & 1.343 & 0.081 & 216 \\ [0.5em]
  \hline 
  \citeauthor{coleman1995} & 1.175 & 0.980 & 1.378 & 1.502 & 2760 & 1.355 & 0.082 & 222 \\ [0.5em]
 \bottomrule
\end{tabular}
\end{center}
\end{table}
Figure ~\ref{fig:channel_mean}(b) shows the mean velocity distribution in wall variables; $u$ normalized by friction velocity $u_\tau$ and $y$ with $\nu / u_\tau$. Present DNS results compare well with \cite{coleman1995} results.  The results are also compared with incompressible law of the wall for turbulent flows, which states that in the viscous sublayer region ($y^+ ~< ~10$) 
\begin{equation}
 \displaystyle u_{wall} = y^+
\end{equation}
and in the log-law region ($10~<~y^+ ~< ~100$) 
 \begin{equation}
\displaystyle {\langle u \rangle}^+ = \frac{1}{\kappa} \mathrm{ln}~y^+ + B
\end{equation}
where $\kappa ~=~ 0.41$ is the Von K\'arm\'an constant and $B ~=~ 5.2$ is the intercept at $\textrm{ln} y^+ = 0$.

In both ANUROOP and \cite{coleman1995} simulations, ${\langle u \rangle}^+$ agrees well with the law of the wall in the viscous sublayer, however it is far away in the log-law region. For high mach number flows, where density variation is very high, variables are not expected to follow the incompressible laws.  However a density-weighted transformation of mean velocity (also known as the Van Driest transformation),  can be used to enable this comparison. This transformation is given as: 
 \begin{equation}
 \displaystyle {\langle u \rangle}_{VD}^+ = \int_0^{{\langle u \rangle}^+} {\bigg(\frac{\langle \rho \rangle}{\rho_w}\bigg)}^\frac{1}{2} d{\langle u \rangle}^+
 \end{equation}
 This transformed velocity is expected to satisfy the incompressible log law (\cite{bradshaw1977compressible})
 \begin{equation}
\displaystyle  {\langle u \rangle}_{VD}^+ = \frac{1}{\kappa} \textrm{ln}~y^+ + B
\end{equation}

The above law is also plotted in Fig.~\ref{fig:channel_mean}(b) and it can be noted that this transformation brings the profile closer to the incompressible log-law; however in the overlap region (\(5 ~< ~y^+~ < ~10 \)) the agreement becomes worse. ~\cite{wei2011direct} argue that a power law is slightly better and less dependent on Mach number than the log-law in this region. 

\def\figsubcap#1{\par\noindent\centering\footnotesize(#1)}
\begin{figure}
  \hspace*{4pt}
    \parbox{2.5in}{\includegraphics[scale=0.25]{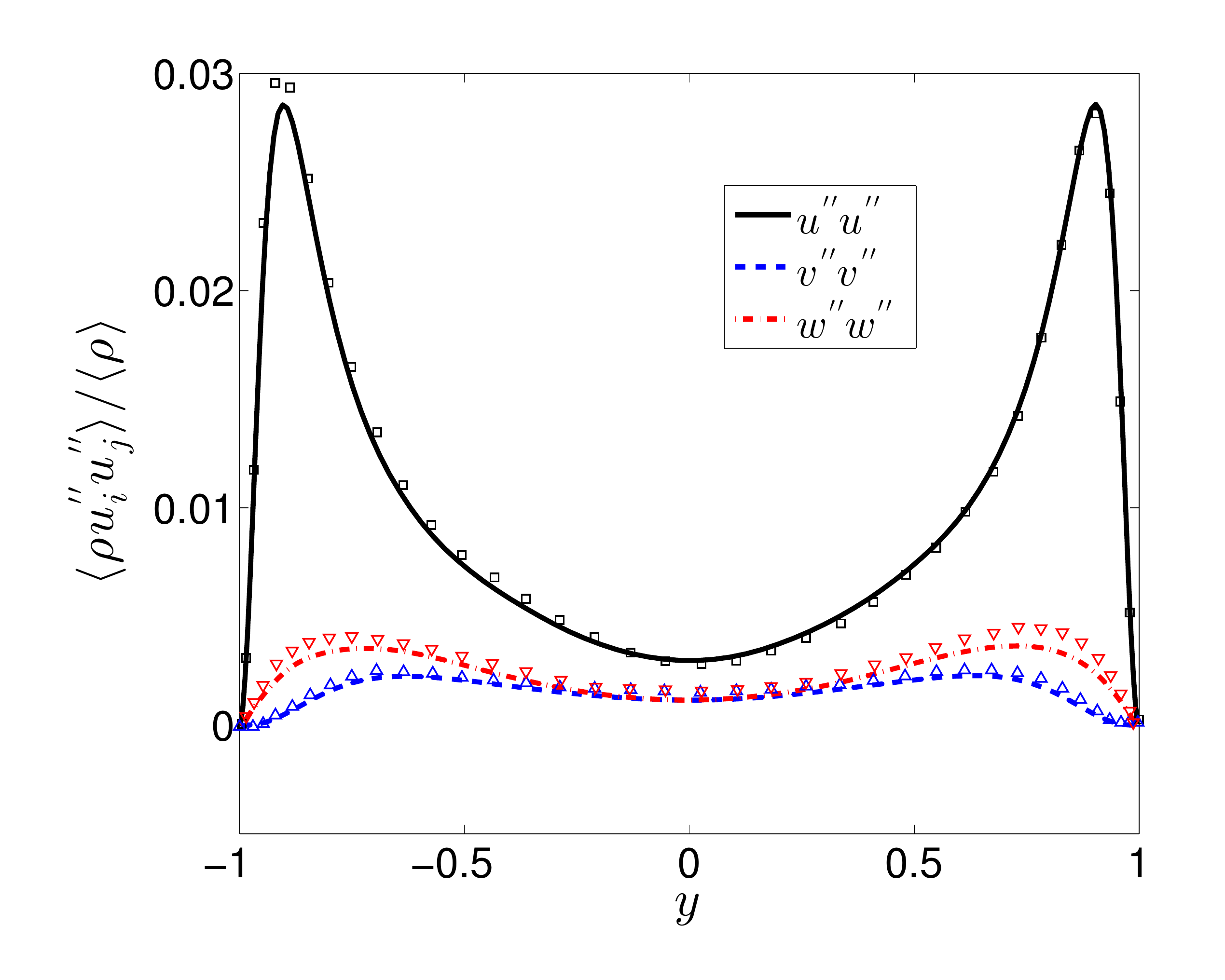}\figsubcap{a}}
  \hspace*{4pt}
    \parbox{2.3in}{\includegraphics[scale=0.22]{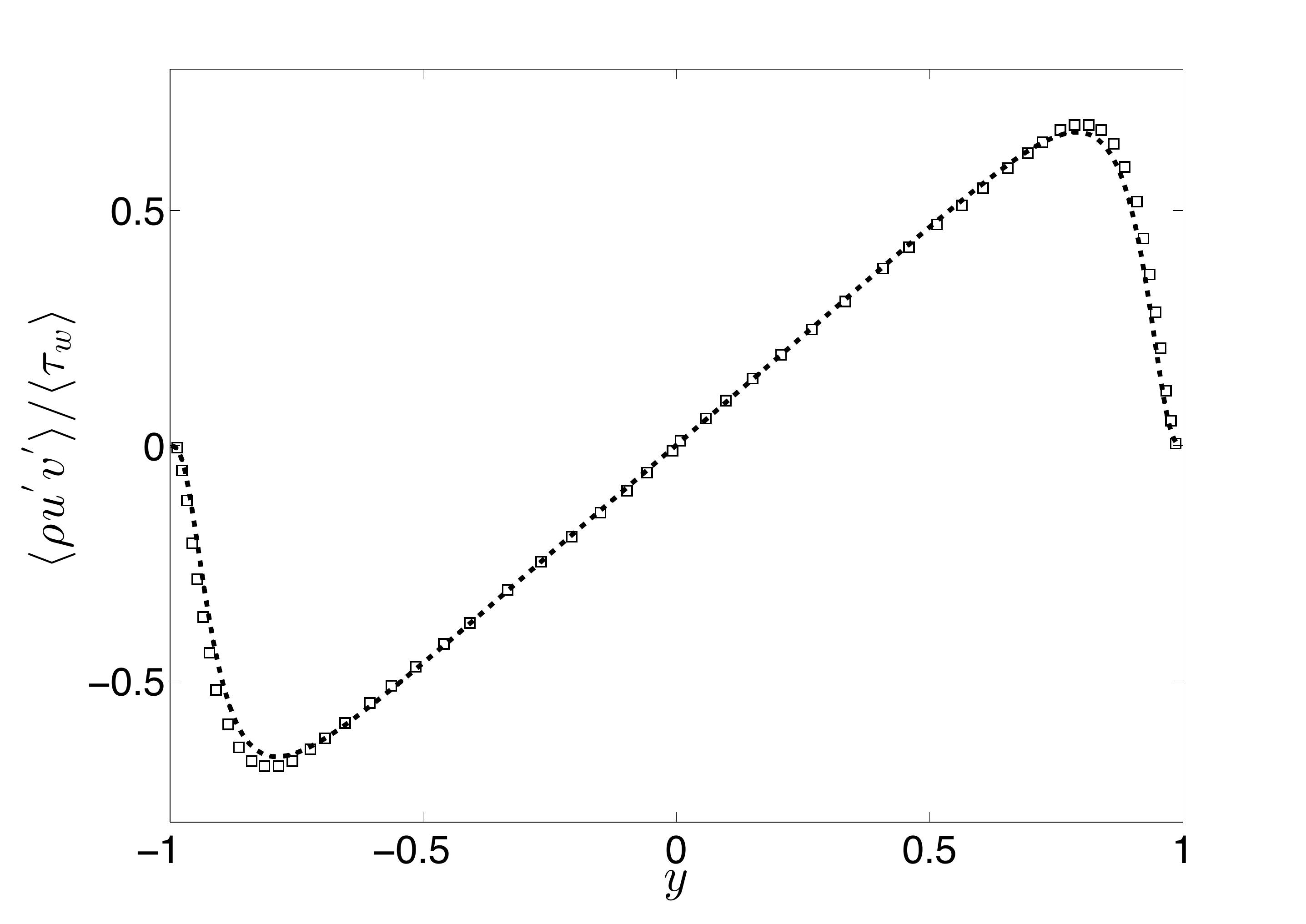}\figsubcap{b}}
  \caption{(a) Turbulent normal stresses ($\langle \rho u^\prime u^\prime \rangle$). (b)Reynolds stress normalized by the wall-shear stress. Symbols are values from \cite{coleman1995} simulation}
\label{fig:channel_rms}
\end{figure}

\subsubsection*{Higher Order Statistics}
Higher order statistics are obtained using the Favre-averaged fluctations. Figure ~\ref{fig:channel_rms}(a) compares the mean turbulent normal stresses  $\langle \rho u^\prime u^\prime \rangle$ for 3-velocity components  with \cite{coleman1995}. Reynolds stress normalized by wall-shear stress is plotted in Fig. \ref{fig:channel_rms}(b). Both these plots show a very good match with the \cite{coleman1995} results. This confirms the ability of ANUROOP to accurately predict the higher-order statistics for a supersonic wall-bounded turbulent flow. 

\part{LPT Blade Simulation}

\section{Flow past T106A blade}
The validated DNS code ANUROOP is used to study flow past a high lift low pressure turbine (LPT) blade T106A in a cascade (Fig. \ref{fig:grid_blade}(a)). The experiments reported by  \cite{stadtmuller1} provide the flow parameters used in the present DNS study. The tests were performed in the High Speed Cascade Wind Tunnel of the Universit{\"a}t der Bundeswehr, M{\"u}nchen, Germany, with the purpose of providing test data at low $Re$ suitable for  DNS studies.  The surface pressure measurements and hot-film traces clearly indicate a separation bubble near the trailing edge (TE) of the suction side; however it is not clear whether the bubble reattaches. The report does not mention the presence of any separation bubble near the leading edge (LE) as found in many DNS studies at low resolution (\cite{kalitzin2003dns, wissink2003, wissink2006, michelassi2002, rajesh2015IUTAM}).   

All the measurements needed for the flow condition are taken in the outflow plane of the cascade, and isentropic relations are used to get the upstream conditions for the computational studies. We use $Re = 51,831$ based on the inlet velocity and the axial chord length ($l_{ax}$) as used in most of the DNS studies in the past for this experiment. \cite{stadtmuller1} mentions considerable uncertainty over the exact inlet angle to be chosen for the computational study, due to the presence of the wake generator ahead of the cascade. Based on RANS simulations and 3D hot wire measurements, \cite{stadtmuller1}  has estimated the real inlet flow angle as $45.5^{\circ}$. This value is used for the present DNS study.

Further, the experiments were performed in a steady upstream flow (disturbance free environment) as well as with upstream wakes created by a moving-bar wake-generator. We have performed simulations without free-stream turbulence as well as with turbulent intensities of 5\% and 10\%. However, the present discussion confines itself to computational study with steady-state measurements (without any free-stream turbulence or wake) as the focus is on a detailed boundary layer study with minimum uncertainties. Results with more complex inlet conditions will appear in later studies.  

 \subsection{Computational domain and grid strategy} 

The computational domain, illustrated in Fig. \ref{fig:grid_blade}, covers the flow around a blade with periodic boundary conditions that simulate the row of blades at top and bottom. The outflow plane is located at a distance of one axial chord $l_{ax}$ downstream of the TE, and the inflow plane is at a distance of half the axial chord upstream of the LE. The pitch between the blades is $p = 0.9306~l_{ax}$, and periodic boundary conditions are applied everywhere on the boundary in the pitchwise $y$-direction as well as spanwise $z$-direction. A no-slip isothermal boundary condition is applied at the blade surface.

The elliptic scheme used in earlier work (\cite{wu2001, kalitzin2003dns, wissink2003, wissink2006, michelassi2002}) offers little control over grid generation in the interior. \cite{kalitzin2003dns} have reported a large number of skewed elements in the passage near the TE  using this scheme, due to the strong curvature and requirement of periodicity in the pitchwise direction. Their results near the LE also exhibit the effect of singularity in the forced $H$-mesh used in the study. \cite{wissink2003} has reported the effect of streamwise resolution on the accuracy with which the separation bubbles can be resolved. \cite{garai2015dns} have reported oscillations in their simulations in the pressure distribution due to low-order geometry representation of the blade profile. These oscillations become more prominent in the high curvature regions. 

Because of these observations and the present emphasis on the boundary layer, greater control and flexibility on the grid in the boundary layer was sought  on the following three aspects:

\begin{itemize}
\item Orthogonal mesh without any singularity near the blade wall 
\item Control on boundary layer grid with respect to the first grid-point near the wall and the ratio of heights of the successive layers
\item Control on streamwise resolution without unduly increasing the overall grid size 
\end{itemize}

\def\figsubcap#1{\par\noindent\centering\footnotesize(#1)}
\begin{figure}
  \hspace*{4pt}
    \parbox{2.4in}{\includegraphics[scale=0.25]{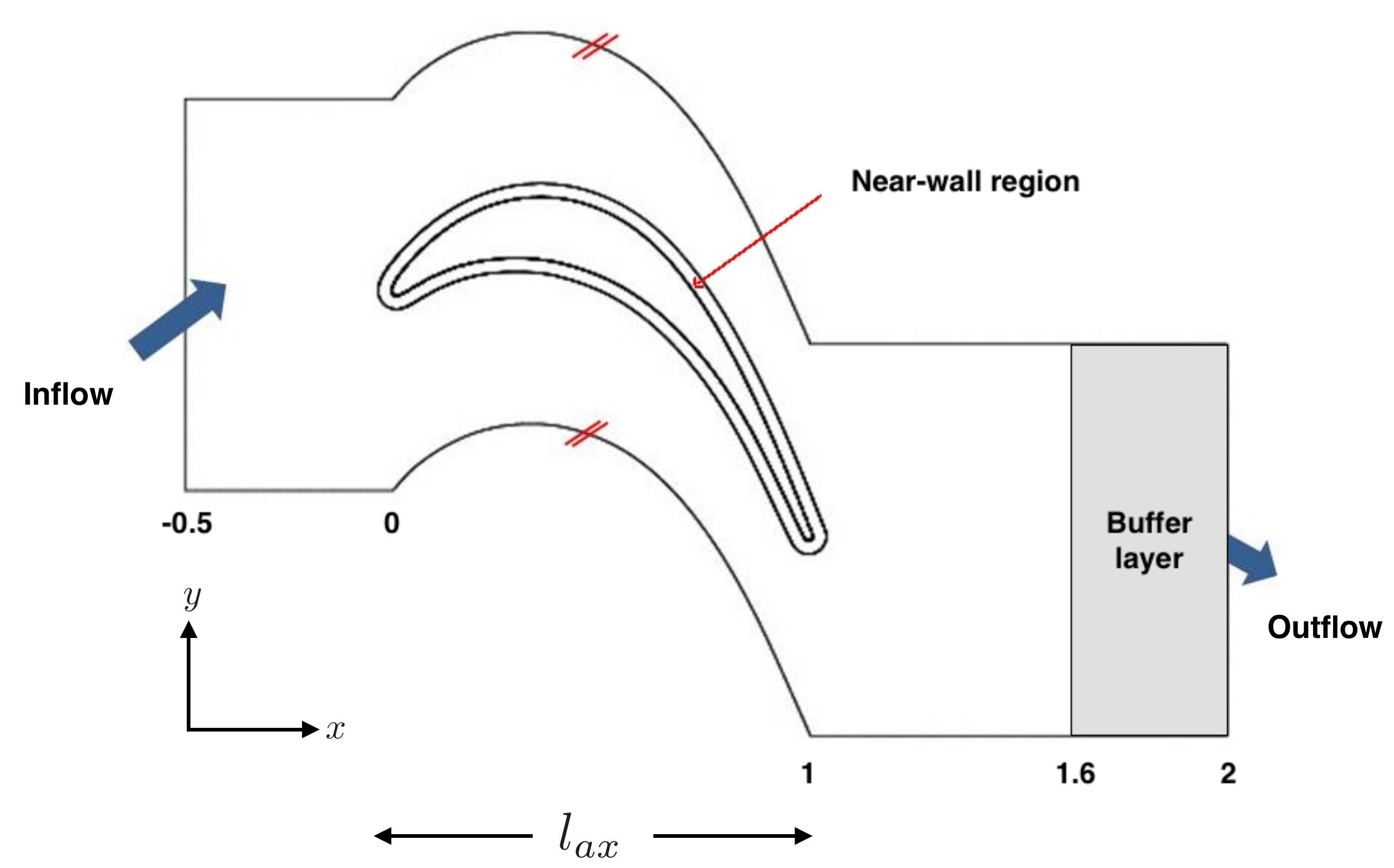}\figsubcap{a}}
  \hspace*{4pt}
    \parbox{2.4in}{\includegraphics[scale=0.25]{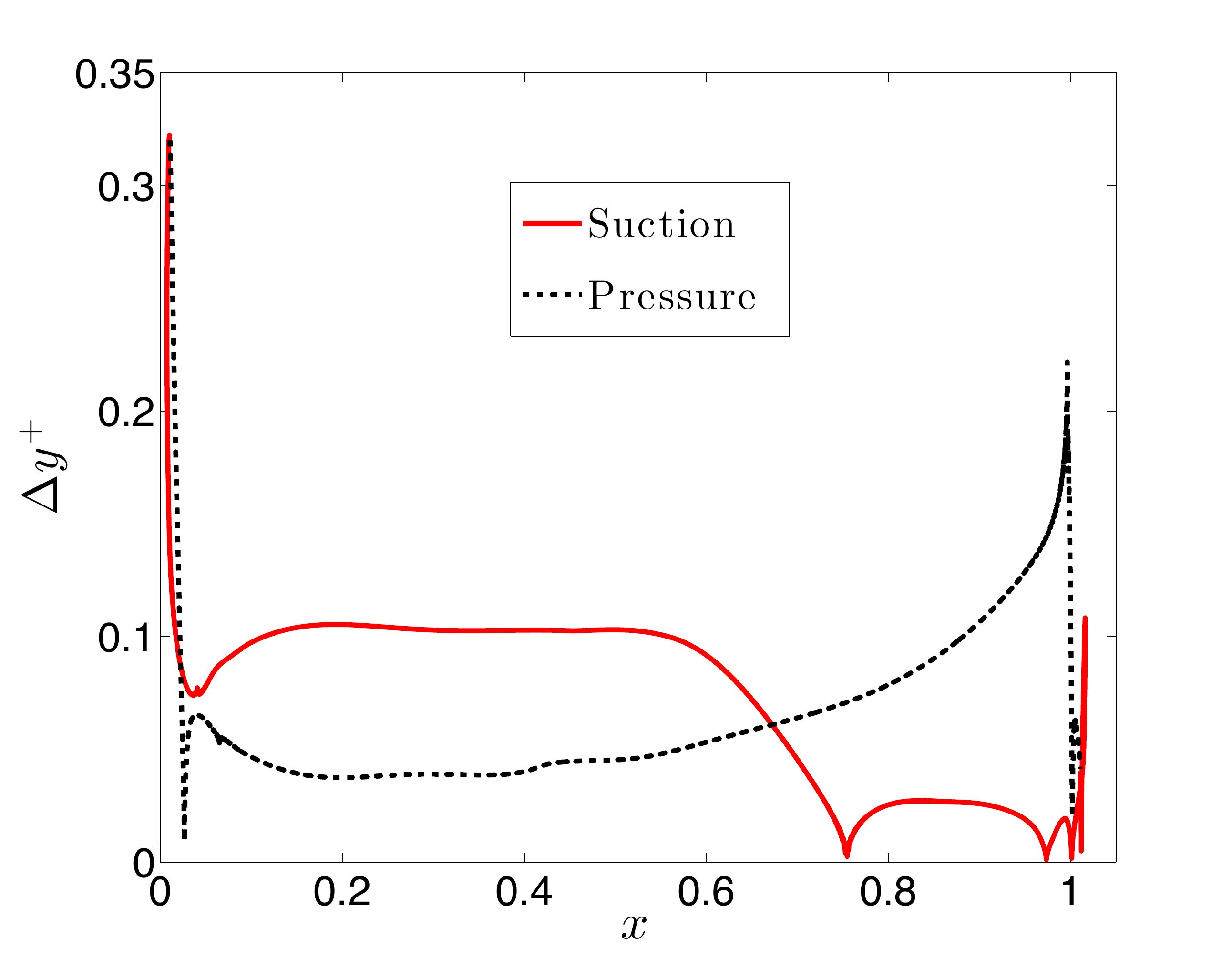}\figsubcap{b}}
  \caption{(a) Computational domain and Grid strategy. (b) $y^+$ distribution on the blade}
    \label{fig:grid_blade}
\end{figure}

It is difficult to meet all these requirements in the elliptic mesh approach, especially because of the periodicity of the mesh elements in the pitchwise ($y$) direction. In our approach, the domain has been divided into two regions: `near-wall' and `outer' (Fig. \ref{fig:grid_blade}).  In the near-wall region,  an orthogonal boundary layer mesh is built by specifying the first grid point away from the wall, the successive ratio and the number of rows in the boundary layer.  After the near-wall mesh is ready, the outer region was filled with quadrilateral elements ensuring proper mesh distribution along the rest of the edges of the domain.  Periodicity is enforced in the pitchwise $y$-direction by exactly copying elements along the periodic edges. It requires a few iterations to get the desirable mesh with no abrupt jump and of good quality. More than 50\% of  the total number of mesh elements are confined in the near-wall region. 

In the streamwise direction, the blade airfoil was divided into 3 parts on both suction and pressure sides, and a very fine clustered grid is used in the high curvature regions (near the leading and trailing edges) for good representation. The total number of grid points along the surface of the blade is 4140, which is an order of magnitude higher than in the most recent simulation (\cite{michelassi2015compressible}). In the `near-wall' mesh, the boundary layer is resolved with 200 layers with the first layer at a height of $9.3 \times 10^{-5} l_{ax}$ (0.1 in wall units) and nearly constant spacing between successive layers.  The distribution of first layer height in wall units thus obtained using simulation results  is shown in Figure ~\ref{fig:grid_blade}. The grid refinement in the boundary layer ensures that the first layer distance in wall units $\Delta y^+$ is below 0.1 throughout the blade for this simulation. 

In the spanwise direction, 128 intervals have been used; thus, the overall number of grid points of the DNS is $161\times10^6$. The highest resolution used todate for the present $Re$ was 17 million \citep{michelassi2002, wissink2003} for the incompressible studies and 18.1 million \citep{michelassi2015compressible} for a recent compressible study. Table ~\ref{dns_lpt} lists the DNS performed on this blade with the grid size as available in the literature. The need for the high resolution will become clear from the comparisons presented in section \ref{sec:T106A_results}.  

\begin{table}
\caption{Grid used for DNS of flow past LPT blades}
\begin{center}
\label{dns_lpt}
\centering 
\def\arraystretch{1.5}
\begin{tabular}{l l l l l l l}

\toprule
\multirow{2}{*}{\bfseries{Ref.}} & \multirow{2}{*}{$\mathbf{Re}$} & \multicolumn{5}{c}{\bfseries{Grid details}}\\
\cline{3-7}
& & $N_x \times N_y \times N_z $& \bfseries{Size} & $\Delta x^+$ & $\Delta y^+$ & $\Delta z^+$ \\
\midrule
\cite{wu2001}  & $1.48\times10^5$ & $1152\times384\times128$ &   $57$ & - & -& -\\

\cite{wissink2003}  &  51,800 & $1014\times266\times64$  & $17$  & 10 &0.8 &3 \\

\cite{kalitzin2003dns} & $1.48\times10^5$ & $1152\times576\times128$  & $85$  & 28 &2.3&1.9\\

\cite{wissink2006} & 51,831 & $1014\times266\times64$ & $17$  & -&-&-\\

\cite{wissink2006direct} & 51,800 & $1014\times260\times64$ &  $17$ & -&-&-\\

\cite{michelassi2015compressible} & 59,634\footnote{based on exit velocity} & $274,176\times66$  & $18$  & 10 &1.4&11\\

ANUROOP & 51,831 & $1,257,162\times128$  & $160$  & 1.1 &0.1&2\\
\bottomrule
\end{tabular}
\end{center}
\end{table}

\section{Results and discussion} \label{sec:T106A_results}
The simulation was performed using the supercomputing facilities available  at  CSIR-4PI, Bangalore.  The time step employed is of the order of $10^{-5} l_{ax} / U$ corresponding to $\mathrm{CFL} ~\approx~ 1.0$, where $U$ is the inlet velocity.  With this step-size, one simulation takes around 10 days with 1024 Intel Xeon E5-2670 (Sandybridge) 2.6 GHz processors ($\sim$ 22 peak TFLOPs) for the flow to progress 10 flow-times. To the best of the authors' knowledge, the present simulation for this set-up is the most compute intensive simulation, with the highest grid resolution till date.

 In the simulations, the viscosity is gradually increased to 10 times of the physical viscosity in the buffer region defined between $1.6 \le x \le 2$ (see  Fig. ~\ref{fig:grid_blade}) to ensure smooth outflow.

Before performing simulations with the present grid, a grid convergence study had been performed for the present set-up with different grid sizes.  The results of this exercise are briefly summarized in \cite{rajesh2015IUTAM}.  In the following description, only results with the grid  described in the preceding section are reported. The results on the pressure distribution and shear stress are reported and brief comments about the separation bubble are given.  As it is found that the longitudinal curvature for this high-lift blade plays a major role in deciding the boundary layer profile, a detailed study of the curvature effect is also carried out. 

\subsection{Pressure distribution} 
In the DNS studies of \cite{michelassi2002}, \cite{wissink2003} and \cite{wissink2006}, there is a significant deviation in the prediction of pressure on the suction side compared to experiments. The reasons given for such discrepeancy are uncertainty in inflow angle and compressibility effects (they solved incompressible Navier-Stokes equations).  The effect of streamwise resolution has been acknowledged in the simulations of \cite{wissink2003}).  

The plot in Fig. ~\ref{fig:blade_cp} shows a comparison of mean static pressure coefficients ($c_p ~ = ~ (p - p_2)/(p_{t1}- p_2)$), where $p_{t1}$ and $p_2$ are total pressure at the inlet and back pressure at the outlet respectively), as obtained by the present simulation against experimental results \citep{stadtmuller1} and also the data from \cite{wissink2003} for the clean inlet (without wake or free-stream turbulence).  $c_p$ is plotted against a streamwise co-ordinate along the arc of the surface of the blade as it is the gradient along arc $s$ that is relevant for the boundary layer on a curved surface. The experimental data as well as data from \cite{wissink2003} are only available along the axial chord ($x$). They are here interpolated on to the streamwise co-ordinates along the surface of the blade and plotted against our results. 

Unlike \cite{wissink2003}, the results with ANUROOP match very well on both sides. This eliminates the doubts about uncertainty in the inflow angle as far as the comparison among simulations is concerned. However, it is not certain if compressibility affects the pressure distribution significantly, as the current simulation has also been conducted at a fairly low mach number (maximum $M$=0.3). In any case, the grid resolution, both in the streamwise as well as  the wall-normal direction, seem to affect the solution significantly, as is also reported in  ~\cite{rajesh2015IUTAM} and \cite{garai2015dns}. 

\subsection{Separation bubble}
\cite{curtis1997development} have shown that for a conventional low pressure turbine blade, 60\% of the losses of a given profile is due to the suction surface boundary layer while the relatively calm pressure side contributes little to the losses. \cite{howell2001high} further showed that out of this 60\%, the presence of separation bubbles generates 60\%. Hence there is a lot of emphasis on identification and classification of separation bubble in the turbine literature. 

\begin{figure}
\centering
\includegraphics[trim=0 0 0 0, clip, width=0.8\linewidth, angle=0]{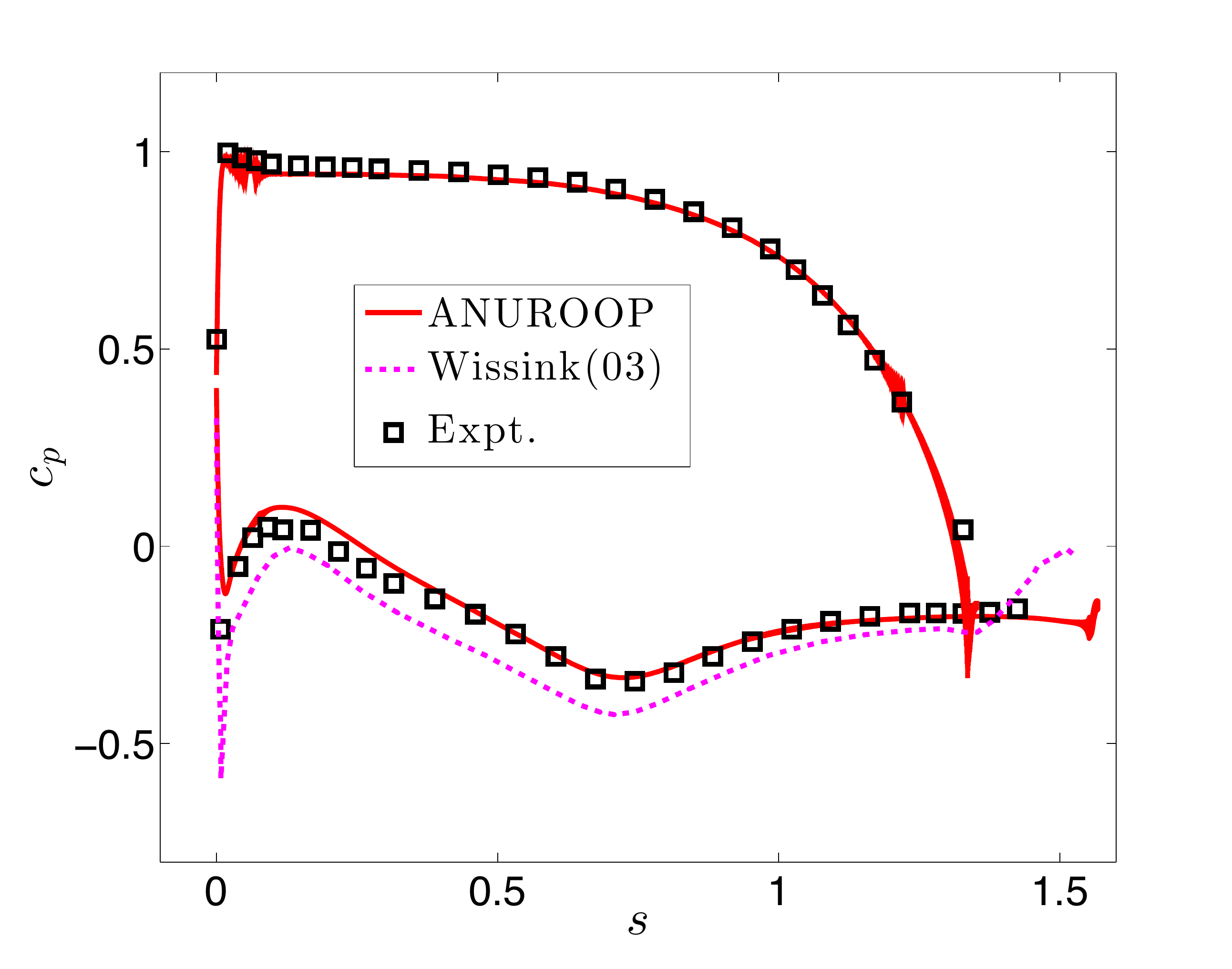}
\caption{$C_p$ along the chord length. Symbols are values from \cite{stadtmuller1} experiments}
\label{fig:blade_cp}
\end{figure}

\begin{figure}
\centering
\includegraphics[trim=0 0 0 0, clip, width=1.0\linewidth, angle=0]{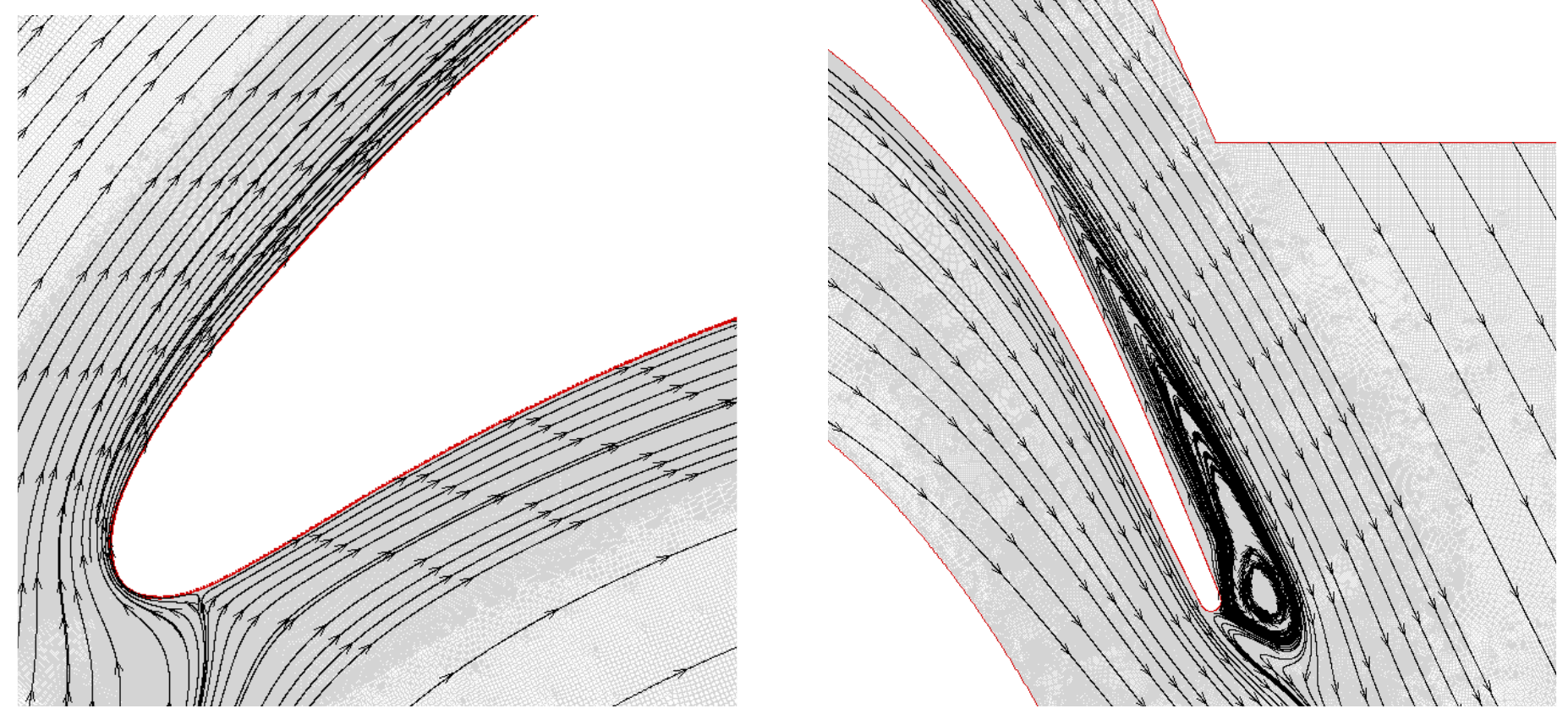}
\caption{Streamlines of mean flow. Left: LE, Right: TE}
\label{fig:separation_all_grids}
\end{figure}

\begin{figure}[h!]
\centering
\includegraphics[trim=0 0 0 0, clip, width=0.8\linewidth, angle=0]{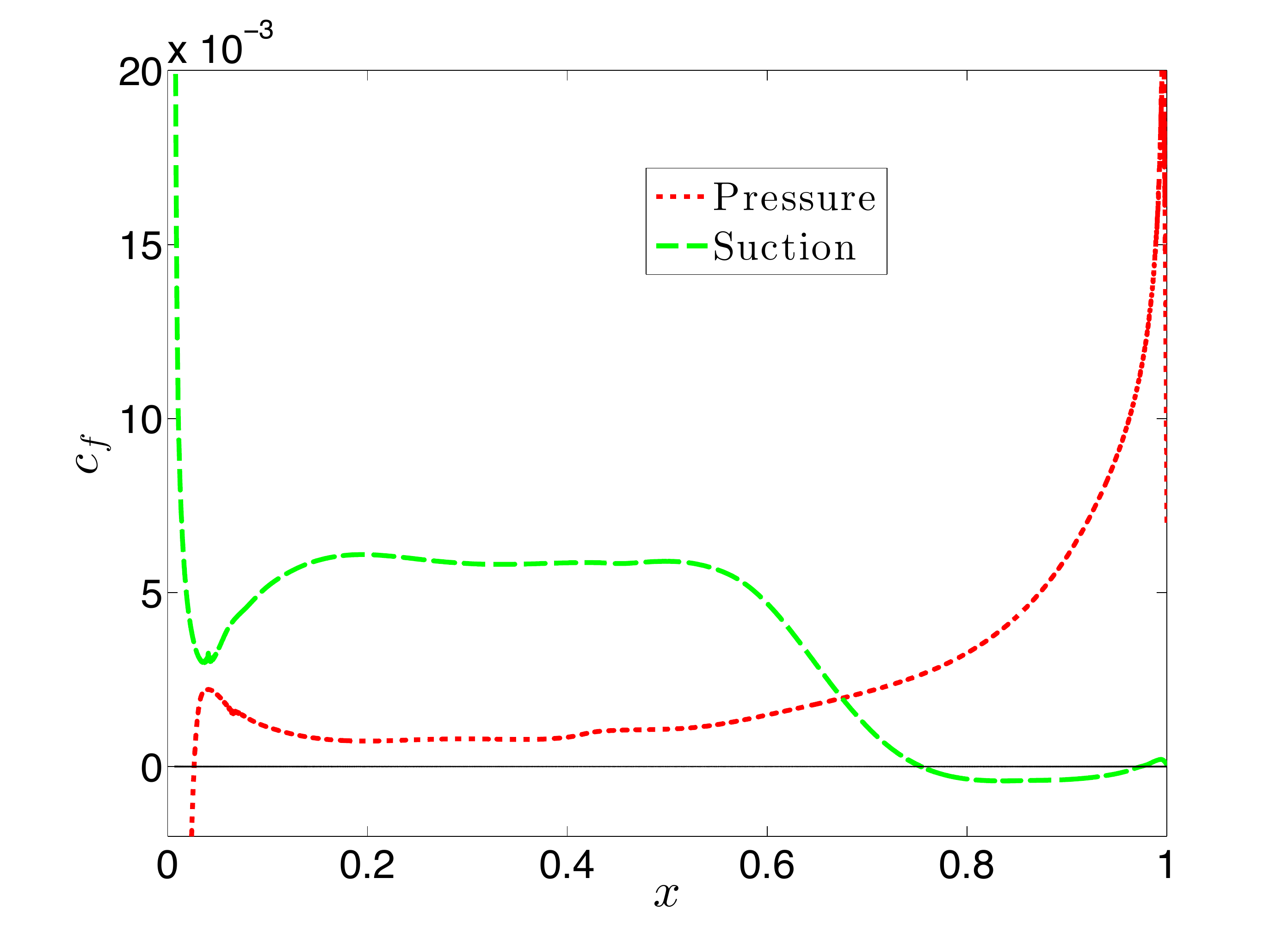}
 \caption{Mean skin-friction on the blade.}
 \label{fig:blade_cf}
\end{figure}

Figure ~\ref{fig:separation_all_grids} gives a visual representation of this bubble through a streamline plot for the current simulation. The bubble begins at $x = 0.76$ and extends upto $x = 1.0$. The long plateau in  the $c_p$ plot (Fig. \ref{fig:blade_cp}) near the trailing edge, also confirms the presence of separation bubble in the experiment as well as the present DNS. The mean skin friction coefficient (wall shear stress normalized by the far free-stream quantities, $\displaystyle c_f = \tau_{w}/(p_{t1}- p_2)$)  from the simulations is plotted in Fig. \ref{fig:blade_cf}. The relatively large value of $c_f$ indicates that the flow is laminar on most of the suction side till separation occurs. The trailing edge separation on the suction side is marked by negative values of $c_f$.  The pressure side shows no signs of separation.

The separation bubbles are often classified as `long' or `short', in order to assess their effect on losses. Long bubbles (usually associated with large losses) are those which affect the pressure distribution throughout the blade. Short bubbles, on the other hand, have only local effect on the distribution. A more precise criterion for characterizing bubbles as of the `long' or `short' type was suggested by \cite{diwan2006bursting}. According to their survey of many separating flows,  bubbles fall in the category of short if  $\displaystyle P \equiv (h^2/\nu)(\Delta u/\Delta x) > -28$, and long otherwise.  Here $h$ represents the maximum thickness of the bubble and $\displaystyle \Delta u/\Delta x$ is a measure of velocity gradient across the bubble. For the present DNS, the value of $P$ is around -120, which confirms that this bubble comes under long category.  This is also confirmed by the large effect of the bubble on the pressure distribution on the blade surface near the trailing edge ($1<s<1.4$). The presence of long bubble is not unexpected as the underlying $Re$ is only $0.5\times10^5$ \citep{hodson2005role}. The size of separation bubbles, and the losses due to them, decrease if higher levels of free-stream turbulence or incoming wakes are introduced \citep{howell2001high}; the bubble may sometimes even disappear. 

In the simulations of \cite{wissink2003}, however, a leading edge bubble is also present apart from the  trailing edge bubble. The trailing edge bubble in their case, however, can be categorized in the `short' category as only a small kink can be noticed in the pressure distribution (Fig. \ref{fig:blade_cp}). It is to be noted that no LE separation is reported by \cite{stadtmuller1} in his experiments. On the other hand, \cite{michelassi2002} argued that the LE separation, if present, might have been missed due to insufficient resolution of the measurement points. 

\subsection{Role of surface curvature} 
Surface streamline curvature plays an important role in determining the nature of the boundary layer flow on the turbine blades.On the convex side of the blade, the flow is stabilized due to the centrifugal force and thus the turbulence  in the boundary layer gets weaker. On the other hand, turbulence in the boundary layer gets enhanced on the concave side due to curvature because of the G{\"o}rtler instability. ~\cite{garai2015dns} have commented on the  effect of curvature on the solution, and emphasized the need for using a high-order geometry representation of the blade profile in order to get non-osciallatory solutions. 

The present DNS has been carried out with a high-order representation of geometry using the non-uniform rational basis spline (NURBS) and a streamwise resolution approximately 4 times higher than in the other studies mentioned in Table \ref{dns_lpt} .Thus the study of the effect of curvature on the boundary layer solutions, even laminar ones, will be of interest. In the following description, we shall discuss curvature effects on the laminar boundary layer near the leading edge in the light of higher-order boundary layer theory.   

In order to estimate the effect of surface curvature, we first calculate the local geometric curvature $K_{geo}$  using the formula 
 \[
  \displaystyle K_{geo}(x) = \frac{x_s'y_s''-y_s'x_s''}{(x_s'^2+y_s'^2)^{3/2}}
 \]
 where $x_s' = \mathrm{d}x_s/\mathrm{d}s$ etc., which are calculated by fitting arcs over every 3 points using the least-square method. The curvature is non-dimensionalized with blade-axial chord length ($l_{ax}$). 

This curvature on the suction side is plotted along the blade arc length and along the chord  in Fig. \ref{fig:curvature_vs_s} (a,b). The highest curvature is $K_{geo}~=~120$ at $s\sim0$, and drops to $<10$ at $s~=~0.03$. Also, the otherwise smooth variation of the geometric curvature has two kinks at $s~=~0.008$ and $s~=~0.024$. This came as a surprise because the smooth blade curve is made by directly importing the co-ordinates of \cite{stadtmuller1} using NURBS. This curvature is then plotted against the the blade curve (Fig. \ref{fig:curvature_vs_s} (b)). The kinks are visible where the curve changes sign (from concave to convex) and also at the grid location $s~=~0.0235, x~=~0.01$. The presence of kinks suggests rapid changes in the third or higher derivatives at certain points. 
\def\figsubcap#1{\par\noindent\centering\footnotesize(#1)}
\begin{figure}
    \parbox{2.4in}{\includegraphics[scale=0.27]{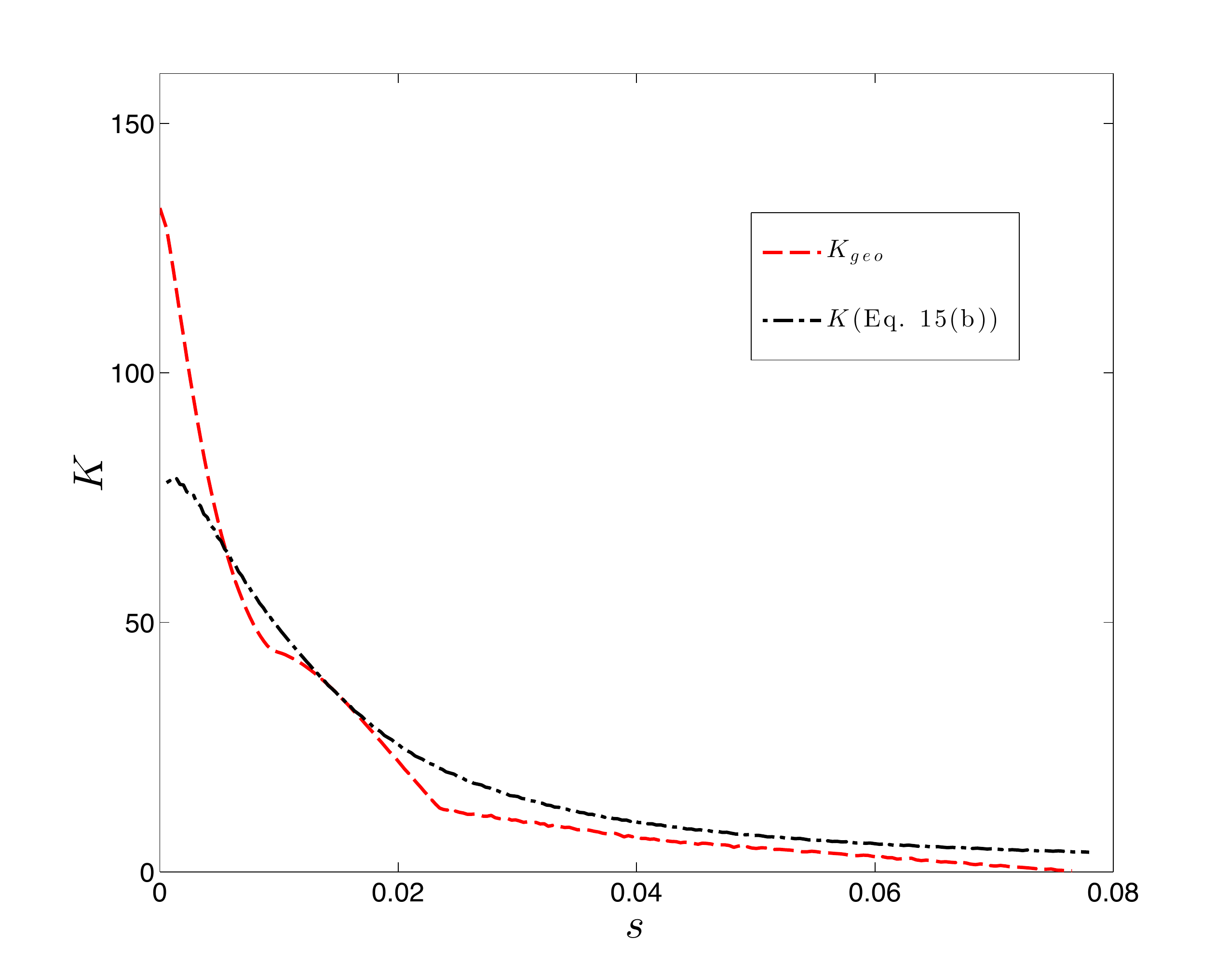}\figsubcap{a}}
  \hspace*{3.2pt}
    \parbox{2.4in}{\includegraphics[scale=0.28]{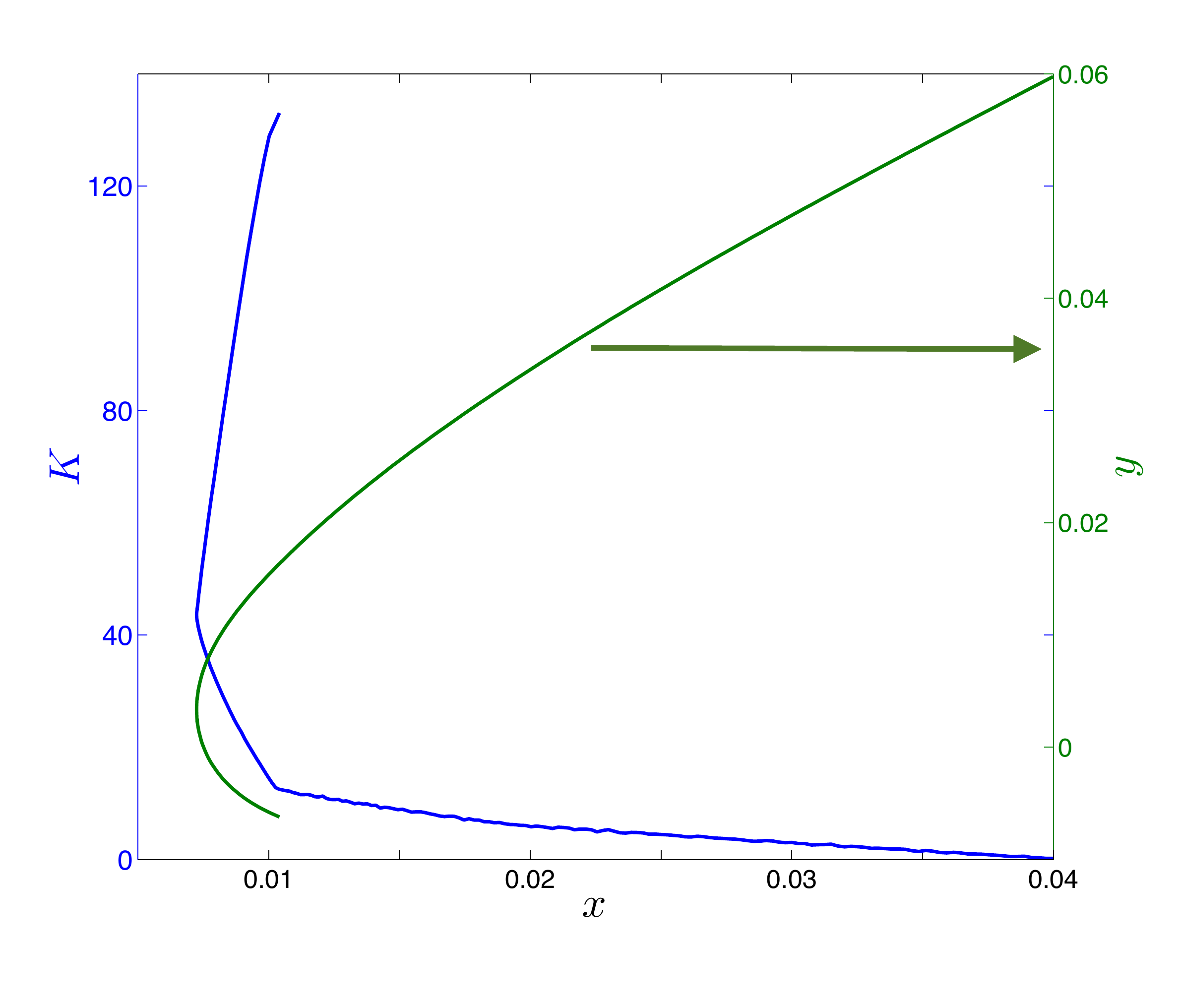}\figsubcap{b}}
  \caption{Curvature along the blade surface (a) and along the axial chord (b) near the leading edge of the suction side}
    \label{fig:curvature_vs_s}
\end{figure}

One of the purposes of DNS simulations is to help improve existing models by providing  new databases and insights. In this spirit, the present DNS results are compared against the higher-order theory of  ~\cite{rnojha1967}. 

The classical boundary layer theory (lowest-order) of Prandtl does not take account of changes due to curvature of the surface on the boundary layer. ~\cite{rnojha1967},  based on higher order  boundary layer theory \citep{van1962higher}, have used inner expansion of variables, such as: $u =  \varepsilon^0 u_0 + \varepsilon^1 u_1 + ...  (\varepsilon = \frac{1}{\sqrt{Re}})$. The $\mathcal{O}(1)$ gives the Prandtl boundary layer equations, where the next order ($\mathcal{O}(\epsilon)$) gives the following $1^{st}$-order (often called $2^{nd}$ order) equations that include curvature effects:

\begin{subequations}
 \label{eqn:first_order}
  \begin{align}
  \displaystyle  \frac{\partial{u_{1}}}{\partial{x}} +  \frac{\partial}{\partial{y}} ({v_{1}} + K ~y ~u_{0}) &= 0 \\
  \nonumber
\displaystyle u_{0} \frac{\partial{u_{1}}}{\partial{x}} + v_{0}  \frac{\partial{u_{1}}}{\partial{y}} +  u_{1} \frac{\partial{u_{0}}}{\partial{x}} + v_{1}  \frac{\partial{u_{0}}}{\partial{y}} =& 
   -  \frac{\partial}{\partial{x}} \bigg\{K \int_0^y u_0^2 ~dy + K \int_0^{\infty} (U_{0s}^2 - u_0^2) ~dy \bigg\}\\ 
   + & \frac{\partial^2 {u_{1}}}{\partial{y^2}} 
   +  K \bigg \{ y~\big(u_{0}\frac{\partial{u_{0}}}{\partial{x}} + \frac{\partial{p_{0}}}{\partial{x}}\big)
   +  \frac{\partial{u_{0}}}{\partial{y}} - u_0 v_0 \bigg \}
\end{align} 
\end{subequations}
with the boundary conditions
 \begin{subequations}
  \label{eqn:first_order_bc}
 \begin{align}
 \displaystyle u_1 &= v_1 = 0, \qquad ~~\mbox{at} \qquad y = 0,\\
 \text{and} \displaystyle \qquad  \qquad u_1 &= - K ~ y ~U_{0s}, \quad \mbox{as} \qquad y \rightarrow \infty.
\end{align}
\end{subequations}
where subscripts 0 and 1 indicate solutions due to Prandtl's boundary layer theory, and the curvature corrections respectively; $K~=~K(s)$ is the curvature of the blade and  $U_{0s} ~=~ U_{0}(X,0)$ is the surface speed from the outer solution as illustrated in Fig. ~\ref{fig:hot_vs_dns2}(b). 

\cite{rnojha1967} have also given a Falkner-Skan like similarity solution for these equations under special conditions, where the velocity varies as a power law with distance as $U_{0s} = Cs^m$. Hence in the transformed co-ordinate ($\eta$) system, the final solution can be  written as $f = f_0 + \varepsilon f_1$. Here $f_0$ is the solution of the usual Falkner-Skan similarity equations,
\begin{align}
\label{eqn:falkner}
 \displaystyle f_0^{'''} + f_0 f_0^{''} = \beta (f_0^{'2} - 1) 
 \end{align}
 where $\displaystyle  \beta = \frac{2m}{m+1}$,
 and the equation for first-order variable $f_1$ is given as: 
\begin{align}
\label{eqn:similarity2}
 \displaystyle \nonumber f_1^{'''} + f_0 f_1^{''} - 2 ~ \beta f_0^{'} f_1^{'} + f_0^{''} f_1 &= \kappa \bigg [ f_0^{''}(\eta f_0 -1) - f_0 f_0^{''} \\
 &- \beta ~ \bigg \{ \eta(f_0^{'2} - 1) - \frac{2}{1+\beta} (f_0^{''} + f_0 f_0^{'} + \beta \eta + A) \bigg\}\bigg]
\end{align}
where
\begin{align}
\displaystyle K &= \kappa {\bigg[ \frac{C(m+1)}{2} \bigg]}^{\frac{1}{2}} s^{\frac{1}{2}(m-1)}
\end{align}
\begin{align}
\displaystyle A &\equiv \lim_{\eta \rightarrow \infty} ~(\eta - f_0)
\end{align}
The boundary conditions in Equations \ref{eqn:first_order_bc} in similarity variables become
\begin{subequations}
\begin{align}
\displaystyle f_1(0) &= 0 = f_1^{'}(0) \\
f_1^{'}(\eta) & \approx -\kappa~\eta \quad \text{as}  \quad \eta \rightarrow \infty
\end{align}
\end{subequations}

Here the primes denote differentiation with respect to the transformed co-ordinate $\eta$. The overall velocity distribution, hence, can be given by $u/U_{0S} = f_0^{\prime} + \kappa \epsilon (f_1^{\prime} / {\kappa})$.

In the present work, both equations \ref{eqn:falkner}  and \ref{eqn:similarity2} were solved sequentially using the shooting method. The Falkner-Skan exponent $m$ was obtained using a local fit for the velocity as shown in Fig. ~\ref{fig:hot_vs_dns2}(a). The effect of curvature can be seen in Fig. ~\ref{fig:hot_vs_dns2}(b) where the velocity profiles with and without curvature effects are compared along with the DNS results for $K = 3.8$ ($s = 0.057$ on the suction side).  The corresponding $K\delta_0^{\star}$ (or equivalently $\kappa \epsilon$), which determines the applicability of $1^{st}$-order theory, is 0.0179 (or 0.01), where $\delta_0^{\star}$ is the boundary layer thickness corresponding to Falkner-Skan solution. The Prandtl boundary layer theory in the figure, shows a significant departure from the DNS solutions, while the $1^{st}$-order theory shows a very good match with the DNS results.
\def\figsubcap#1{\par\noindent\centering\footnotesize(#1)}
\begin{figure}
    \parbox{2.4in}{\includegraphics[scale=0.27]{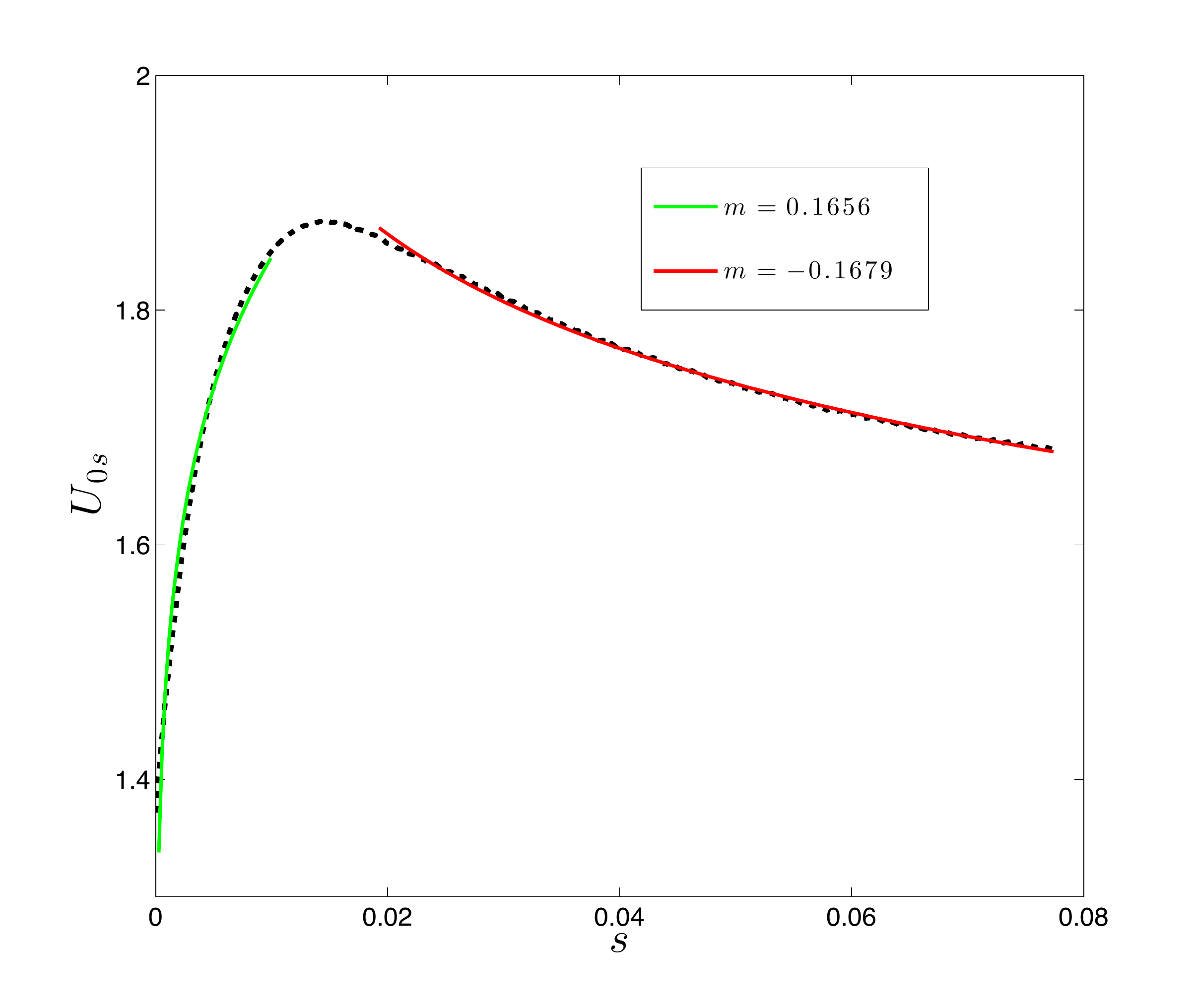}\figsubcap{a}}
    \parbox{2.4in}{\includegraphics[scale=0.27]{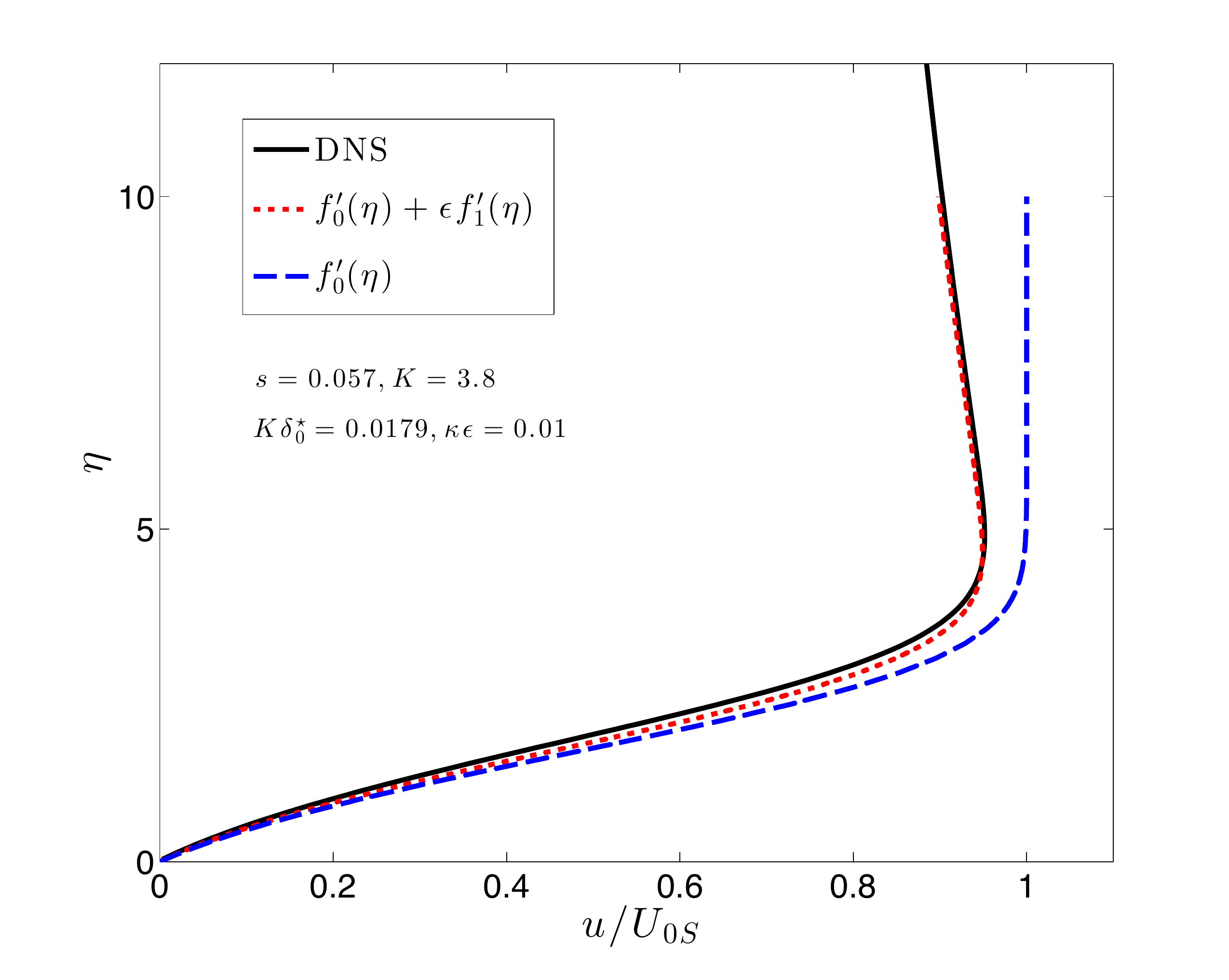}\figsubcap{b}}
  \caption{(a) Surface velocity fit for favourable and adverse pressure gradient (FPG and APG) regions in the locality of high curvature. These fits were used to calculate Falkner-Skan as well $1^{st}$-order solutions (b) Curvature effects: Prandtl vs Higher Order at $s=0.057$. DNS results are also plotted.}
    \label{fig:hot_vs_dns2}
\end{figure}

\begin{center}
 \begin{figure}[t!]
\includegraphics[width=1.0\textwidth]{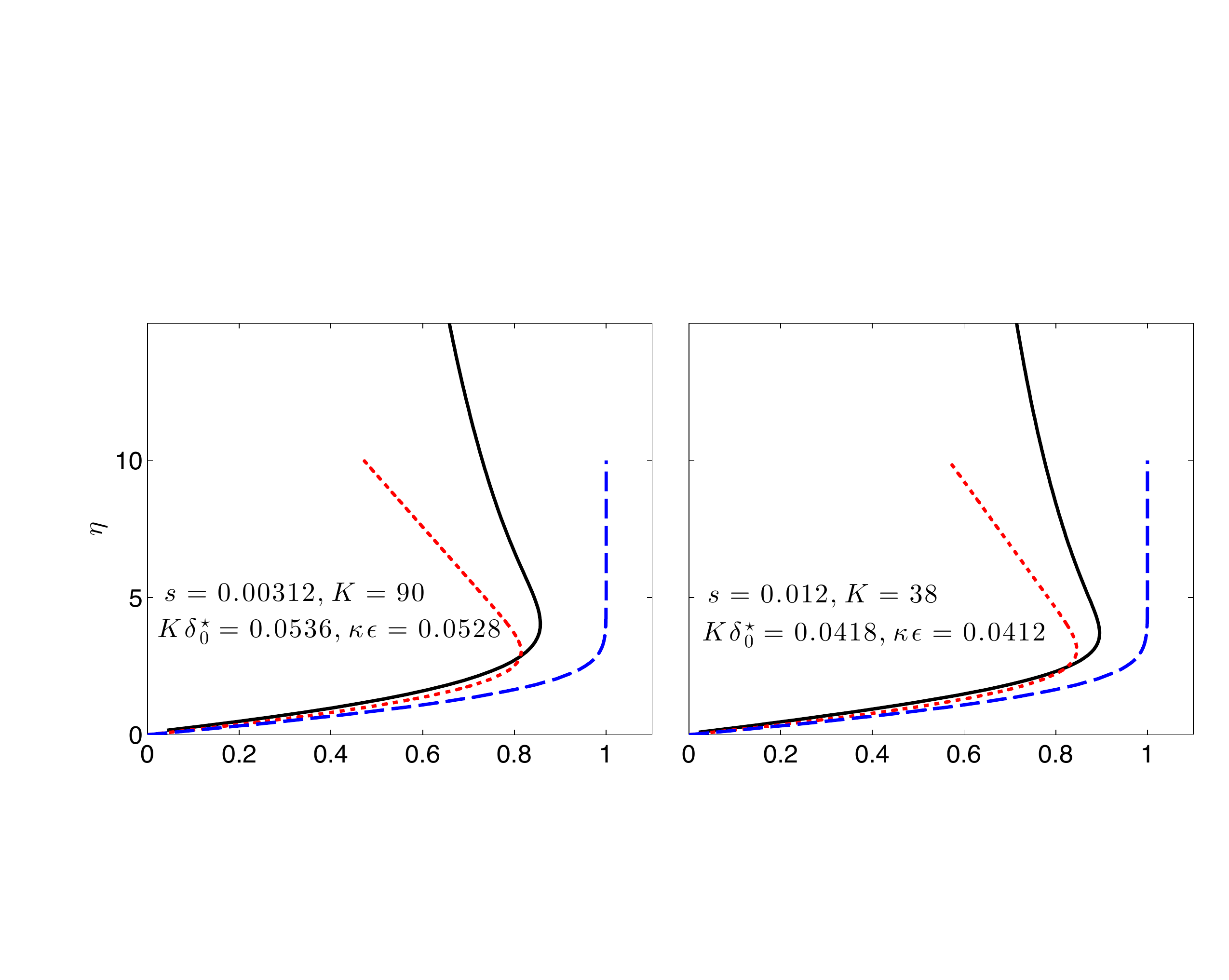}
\includegraphics[width=1.005\textwidth]{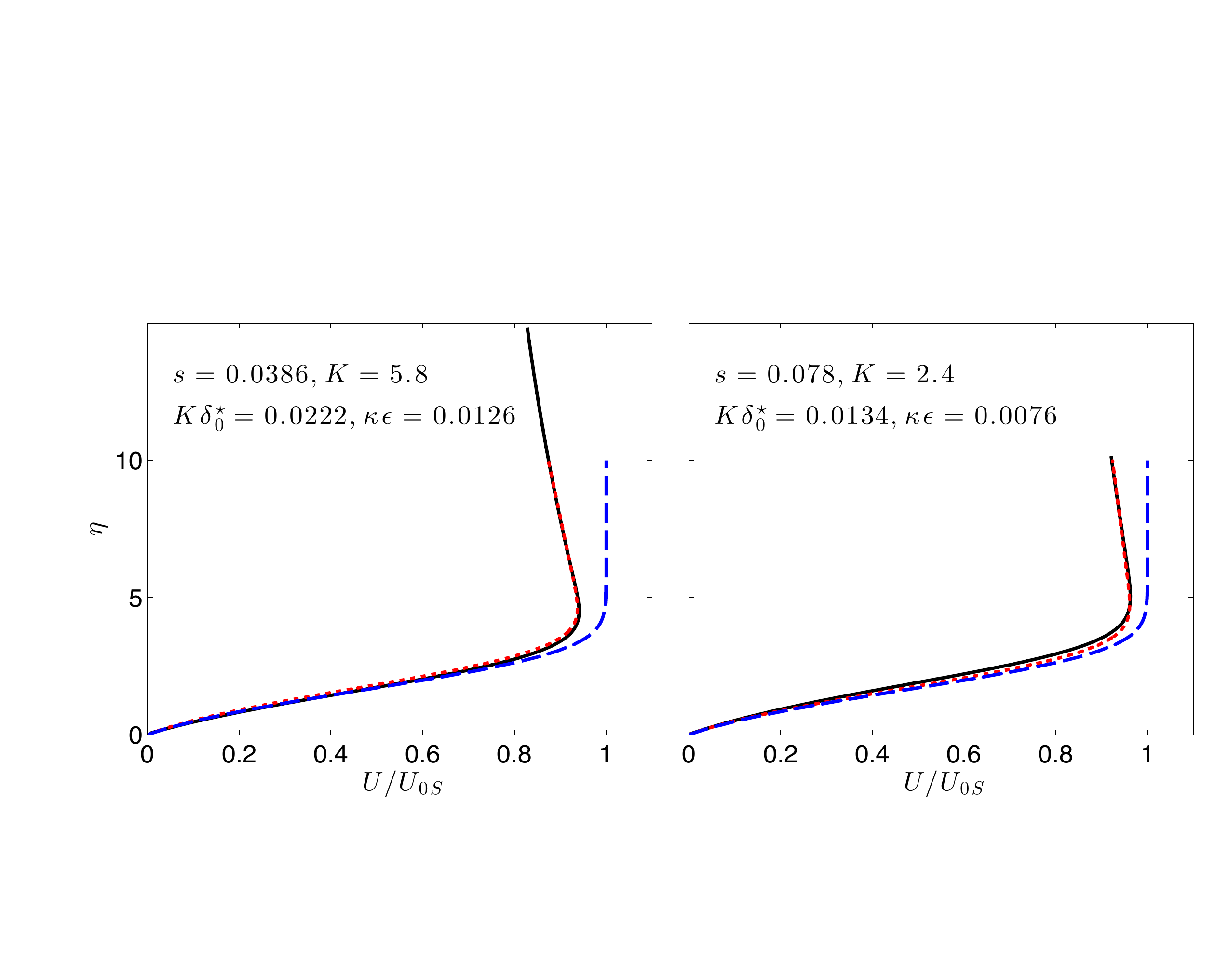}
\caption{Velocity profiles from DNS near the leading edge of the suction side. Legends same as in Fig. \ref{fig:hot_vs_dns2}(b). Top panel: FPG ($m = 0.1656$). Bottom panel: APG ($m = -0.1679$)}
\label{fig:vel_prof}
\end{figure}
\end{center}

One can determine the value of $K\delta_0^{\star}$ at which Equation \ref{eqn:similarity2} fails or separation is suspected. In Fig. ~\ref{fig:vel_prof} we present solutions at four locations. In the top panel, boundary layer profiles in the vicinity of the leading edge are shown, where the pressure gradient is favourable but curvature $K$ is very high ($\mathcal{O}(100)$).  The effect of high surface curvature can be noticed in the `curved'  boundary layer profiles as  the solutions obtained by $1^{st}$-order theory in this region are found to be not adequate.  The region of overlap of both the plots in the outer solution is very small, and one may require even higher order theory in order to predict the boundary layer.

In the bottom panel, results are shown for two stations further downstream, where the pressure gradient is adverse but the curvature is relatively small $[\mathcal{O}(10)]$.  The effect of high surface curvature is weaker in these regions, and $1^{st}$-order  theory seems to be adequate to predict the boundary layer parameters. Similar observations are valid for all stations beyond $s \approx 0.0278$ ($K\delta_0^{\star} = 0.0353$), where the curvatures obtained using the geometric formula as well as $1^{st}$-order theory are comparable (see Fig. \ref{fig:curvature_vs_s}(a)). $1^{st}$-order  theory also predicts that the separation in this region is possible at $m = -0.1679$ if $K\delta_0^{\star} = 0.2624$.

To the best of our knowledge, the present effort is the first assessment of the bounds of validity of the Prandtl boundary layer theory and the next higher order approximation based on DNS results, especially for a problem of practical importance. Though the applicability of above results is restricted to laminar boundary layers, their importance lies in the fact that in many turbine blades the flow at the leading edge remains laminar and is prone to separation. Apart from predicting boundary layer parameters, this theory can be used to predict the separation for such flows. This is of importance to the designer as a separation bubble at the leading edge can trip the boundary layer towards turbulence. The applicability as well as accuracy of this theory can be further improved by directly solving the governing equations instead of using fits to similarity solutions. 

The limitation of this theory is also mentioned above, where the predictions fail in the vicinity of the leading edge for a high lift blade such as T106A. To get more reliable solutions for such flow regimes, DNS may be the only approach which can be deemed adequate.    

\section{Conclusion}
A new DNS code, ANUROOP, has been developed for simulating the flow past a low pressure turbine blade. ANUROOP solves the compressible Navier-Stokes equations in the finite volume formulation and uses a flux-scheme that respects the conservation of evolution of kinetic energy. This code has been validated for its accuracy and robustness for several benchmark problems including  Taylor-Green vortex and supersonic turbulent channel flow. 

ANUROOP was then used to simulate flow past a high lift low-pressure turbine blade T106A at $Re \approx 52000$  and an angle higher than the design incidence. No artificial inflow disturbance was used in the simulation. A highly resolved boundary layer grid was used in all the three directions (streamwise, wall-normal and spanwise), which takes the total mesh count in the computational domain to around 160 million. The no. of mesh points used is approximately one order of magnitude higher than that used in earlier studies for similar flow conditions. As a consequence pressure distribution predicted by the current simulation is closer to the experimental values than in earlier simulations.  This suggests that the earlier explanations of discrepancies in the pressure distributions, in particular those attributing them to experimental uncertainties, may not be justified.

While the flow on the pressure side of the blade remains laminar, a `long' separation bubble forms in the aft region on the suction side disturbing the incoming flow. The  surface curvature of the blade significantly alters the boundary layer, and it is found that the Prandtl boundary layer theory is inadequate to predict such behaviour. The next order boundary layer theory which includes curvature corrections, however, predicts the boundary layer significantly better except in the regions where curvature is very high ($\mathcal{O}(100)$ normalized with axial chord length). 

DNS simulations with free-stream turbulence as well as wake-disturbances, and studies of other boundary layer parameters, will appear in future reports.

\section*{Acknowledgement}
We would like to thank Director, GTRE (Gas Turbine Research Establishment), as well as GATET (Gas Turbine Enabling Technology) for funding this project. Also, thanks are due to Director General, C-DAC, Pune as well as Head, CSIR-4PI for making their compute resources available to us for simulations. Prof. Wolfgang Rodi, Dr. Jan Wissink and Mr. Stephan Stotz are gratefully acknowledged for providing crucial data pertaining to the study.
\section*{References}

\bibliography{refer}

\begin{thebibliography}{36}
\expandafter\ifx\csname natexlab\endcsname\relax\def\natexlab#1{#1}\fi
\providecommand{\url}[1]{\texttt{#1}}
\providecommand{\href}[2]{#2}
\providecommand{\path}[1]{#1}
\providecommand{\DOIprefix}{doi:}
\providecommand{\ArXivprefix}{arXiv:}
\providecommand{\URLprefix}{URL: }
\providecommand{\Pubmedprefix}{pmid:}
\providecommand{\doi}[1]{\href{http://dx.doi.org/#1}{\path{#1}}}
\providecommand{\Pubmed}[1]{\href{pmid:#1}{\path{#1}}}
\providecommand{\bibinfo}[2]{#2}
\ifx\xfnm\undefined \def\xfnm[#1]{\unskip,\space#1}\fi
\bibitem[{Allaneau and Jameson(2009)}]{allaneau2009direct}
\bibinfo{author}{Allaneau\xfnm[ Y.]}, \bibinfo{author}{Jameson\xfnm[ A.]}.
\newblock \bibinfo{title}{Direct numerical simulations of a two-dimensional
  viscous flow in a shocktube using a kinetic energy preserving scheme}.
\newblock \bibinfo{journal}{AIAA paper 2009}
  \bibinfo{year}{2009};\bibinfo{volume}{3797}.
\bibitem[{Allaneau and Jameson(2010)}]{allaneau2010direct}
\bibinfo{author}{Allaneau\xfnm[ Y.]}, \bibinfo{author}{Jameson\xfnm[ A.]}.
\newblock \bibinfo{title}{Direct numerical simulations of plunging airfoils}.
\newblock In: \bibinfo{booktitle}{48th AIAA Aerospace Sciences Meeting
  Including the New Horizons Forum and Aerospace Exposition, AIAA Paper}.
  volume \bibinfo{volume}{728}; \bibinfo{year}{2010}. .
\bibitem[{Blazek(2005)}]{blazek2005computational}
\bibinfo{author}{Blazek\xfnm[ J.]}.
\newblock \bibinfo{title}{Computational Fluid Dynamics: Principles and
  Applications:(Book with accompanying CD)}.
\newblock \bibinfo{publisher}{Elsevier}, \bibinfo{year}{2005}.
\bibitem[{Bradshaw(1977)}]{bradshaw1977compressible}
\bibinfo{author}{Bradshaw\xfnm[ P.]}.
\newblock \bibinfo{title}{Compressible turbulent shear layers}.
\newblock \bibinfo{journal}{Annual Review of Fluid Mechanics}
  \bibinfo{year}{1977};\bibinfo{volume}{9}(\bibinfo{number}{1}):\bibinfo{pages}{33--52}.
\bibitem[{{Coleman} et~al.(1995){Coleman}, {Kim} and {Moser}}]{coleman1995}
\bibinfo{author}{{Coleman}\xfnm[ G.N.]}, \bibinfo{author}{{Kim}\xfnm[ J.]},
  \bibinfo{author}{{Moser}\xfnm[ R.D.]}.
\newblock \bibinfo{title}{{A numerical study of turbulent supersonic
  isothermal-wall channel flow}}.
\newblock \bibinfo{journal}{Journal of Fluid Mechanics}
  \bibinfo{year}{1995};\bibinfo{volume}{305}:\bibinfo{pages}{159--183}.
\newblock \DOIprefix\doi{10.1017/S0022112095004587}.
\bibitem[{Curtis et~al.(1997)Curtis, Hodson, Banieghbal, Denton, Howell and
  Harvey}]{curtis1997development}
\bibinfo{author}{Curtis\xfnm[ E.]}, \bibinfo{author}{Hodson\xfnm[ H.]},
  \bibinfo{author}{Banieghbal\xfnm[ M.]}, \bibinfo{author}{Denton\xfnm[ J.]},
  \bibinfo{author}{Howell\xfnm[ R.]}, \bibinfo{author}{Harvey\xfnm[ N.]}.
\newblock \bibinfo{title}{Development of blade profiles for low-pressure
  turbine applications}.
\newblock \bibinfo{journal}{Journal of Turbomachinery}
  \bibinfo{year}{1997};\bibinfo{volume}{119}(\bibinfo{number}{3}):\bibinfo{pages}{531--538}.
\bibitem[{Diwan et~al.(2006)Diwan, Chetan and Ramesh}]{diwan2006bursting}
\bibinfo{author}{Diwan\xfnm[ S.]}, \bibinfo{author}{Chetan\xfnm[ S.]},
  \bibinfo{author}{Ramesh\xfnm[ O.]}.
\newblock \bibinfo{title}{On the bursting criterion for laminar separation
  bubbles}.
\newblock In: \bibinfo{booktitle}{IUTAM Symposium on Laminar-Turbulent
  Transition}. \bibinfo{organization}{Springer}; \bibinfo{year}{2006}. p.
  \bibinfo{pages}{401--407}.
\bibitem[{Frink(1994)}]{frink1994recent}
\bibinfo{author}{Frink\xfnm[ N.T.]}.
\newblock \bibinfo{title}{Recent progress toward a three-dimensional
  unstructured navier-stokes flow solver}.
\newblock In: \bibinfo{booktitle}{AIAA, Aerospace Sciences Meeting \& Exhibit,
  32 nd, Reno, NV}. \bibinfo{year}{1994}. .
\bibitem[{Garai et~al.(2015)Garai, Diosady, Murman and Madavan}]{garai2015dns}
\bibinfo{author}{Garai\xfnm[ A.]}, \bibinfo{author}{Diosady\xfnm[ L.]},
  \bibinfo{author}{Murman\xfnm[ S.]}, \bibinfo{author}{Madavan\xfnm[ N.]}.
\newblock \bibinfo{title}{{DNS} of flow in a low-pressure turbine cascade using
  a discontinuous-galerkin spectral-element method}.
\newblock In: \bibinfo{booktitle}{ASME Turbo Expo 2015: Turbine Technical
  Conference and Exposition}. \bibinfo{organization}{American Society of
  Mechanical Engineers}; \bibinfo{year}{2015}. p.
  \bibinfo{pages}{V02BT39A023--V02BT39A023}.
\bibitem[{Hillewaert(2012)}]{hiocfd2013}
\bibinfo{author}{Hillewaert\xfnm[ K.]}.
\newblock \bibinfo{title}{Direct numerical simulation of the taylor-green
  vortex at {Re} = 1600, 2nd international high order cfd workshop}.
\newblock \bibinfo{howpublished}{On the WWW. http://www.as.dlr.de/hiocfd/};
  \bibinfo{year}{2012}.
\bibitem[{Hirsch(1988)}]{hirsch1988numerical}
\bibinfo{author}{Hirsch\xfnm[ C.]}.
\newblock \bibinfo{title}{Numerical computation of internal and external
  flows}.
\newblock \bibinfo{journal}{Wiley series in numerical methods in engineering}
  \bibinfo{year}{1988};.
\bibitem[{Hodson and Howell(2005)}]{hodson2005role}
\bibinfo{author}{Hodson\xfnm[ H.P.]}, \bibinfo{author}{Howell\xfnm[ R.J.]}.
\newblock \bibinfo{title}{The role of transition in high-lift low-pressure
  turbines for aeroengines}.
\newblock \bibinfo{journal}{Progress in Aerospace Sciences}
  \bibinfo{year}{2005};\bibinfo{volume}{41}(\bibinfo{number}{6}):\bibinfo{pages}{419--454}.
\bibitem[{Holmes et~al.(1989)Holmes, Connell and Engines}]{holmes1989solution}
\bibinfo{author}{Holmes\xfnm[ D.]}, \bibinfo{author}{Connell\xfnm[ S.]},
  \bibinfo{author}{Engines\xfnm[ G.A.]}.
\newblock \bibinfo{title}{Solution of the 2D Navier-Stokes equations on
  unstructured adaptive grids}.
\newblock \bibinfo{publisher}{American Institute of Aeronautics and
  Astronautics}, \bibinfo{year}{1989}.
\bibitem[{Howell et~al.(2001)Howell, Ramesh, Hodson, Harvey and
  Schulte}]{howell2001high}
\bibinfo{author}{Howell\xfnm[ R.]}, \bibinfo{author}{Ramesh\xfnm[ O.]},
  \bibinfo{author}{Hodson\xfnm[ H.]}, \bibinfo{author}{Harvey\xfnm[ N.]},
  \bibinfo{author}{Schulte\xfnm[ V.]}.
\newblock \bibinfo{title}{High lift and aft-loaded profiles for low-pressure
  turbines}.
\newblock \bibinfo{journal}{Journal of Turbomachinery}
  \bibinfo{year}{2001};\bibinfo{volume}{123}(\bibinfo{number}{2}):\bibinfo{pages}{181--188}.
\bibitem[{Jameson(2008)}]{jameson2008}
\bibinfo{author}{Jameson\xfnm[ A.]}.
\newblock \bibinfo{title}{Formulation of kinetic energy preserving conservative
  schemes for gas dynamics and direct numerical simulation of one-dimensional
  viscous compressible flow in a shock tube using entropy and kinetic energy
  preserving schemes}.
\newblock \bibinfo{journal}{J Sci Comput}
  \bibinfo{year}{2008};\bibinfo{volume}{34}(\bibinfo{number}{2}):\bibinfo{pages}{188--208}.
\newblock \DOIprefix\doi{10.1007/s10915-007-9172-6}.
\bibitem[{Kalitzin et~al.(2003)Kalitzin, Wu and Durbin}]{kalitzin2003dns}
\bibinfo{author}{Kalitzin\xfnm[ G.]}, \bibinfo{author}{Wu\xfnm[ X.]},
  \bibinfo{author}{Durbin\xfnm[ P.A.]}.
\newblock \bibinfo{title}{{DNS} of fully turbulent flow in a lpt passage}.
\newblock \bibinfo{journal}{International Journal of Heat and Fluid Flow}
  \bibinfo{year}{2003};\bibinfo{volume}{24}(\bibinfo{number}{4}):\bibinfo{pages}{636--644}.
\bibitem[{Michelassi et~al.(2015)Michelassi, Chen, Pichler and
  Sandberg}]{michelassi2015compressible}
\bibinfo{author}{Michelassi\xfnm[ V.]}, \bibinfo{author}{Chen\xfnm[ L.W.]},
  \bibinfo{author}{Pichler\xfnm[ R.]}, \bibinfo{author}{Sandberg\xfnm[ R.D.]}.
\newblock \bibinfo{title}{Compressible direct numerical simulation of
  low-pressure turbines---part ii: Effect of inflow disturbances}.
\newblock \bibinfo{journal}{Journal of Turbomachinery}
  \bibinfo{year}{2015};\bibinfo{volume}{137}(\bibinfo{number}{7}):\bibinfo{pages}{071005}.
\bibitem[{Michelassi et~al.(2002)Michelassi, Wissink and Rodi}]{michelassi2002}
\bibinfo{author}{Michelassi\xfnm[ V.]}, \bibinfo{author}{Wissink\xfnm[ J.]},
  \bibinfo{author}{Rodi\xfnm[ W.]}.
\newblock \bibinfo{title}{Analysis of {DNS} and {LES} of flow in a low pressure
  turbine cascade with incoming wakes and comparison with experiments}.
\newblock \bibinfo{journal}{Flow, Turbulence and Combustion}
  \bibinfo{year}{2002};\bibinfo{volume}{69}:\bibinfo{pages}{295--329}.
\newblock \DOIprefix\doi{10.1023/A:1027334303200}.
\bibitem[{Narasimha and Ojha(1967)}]{rnojha1967}
\bibinfo{author}{Narasimha\xfnm[ R.]}, \bibinfo{author}{Ojha\xfnm[ S.]}.
\newblock \bibinfo{title}{Effect of longitudinal surface curvature on boundary
  layers}.
\newblock \bibinfo{journal}{Journal of Fluid Mechanics}
  \bibinfo{year}{1967};\bibinfo{volume}{29}(\bibinfo{number}{01}):\bibinfo{pages}{187--199}.
\bibitem[{Ranjan(2016)}]{rajeshthesis2015}
\bibinfo{author}{Ranjan\xfnm[ R.]}.
\newblock \bibinfo{title}{Compressible DNS studies of the boundary layer on a
  low pressure turbine (LPT) blade at high incidence}.
\newblock Ph.D. thesis; JNCASR; \bibinfo{year}{2016}.
\bibitem[{Ranjan et~al.(2013)Ranjan, Deshpande and
  Narasimha}]{rajesh2013numerical}
\bibinfo{author}{Ranjan\xfnm[ R.]}, \bibinfo{author}{Deshpande\xfnm[ S.]},
  \bibinfo{author}{Narasimha\xfnm[ R.]}.
\newblock \bibinfo{title}{Numerical methodology for simulating flows over
  turbine blades}.
\newblock In: \bibinfo{booktitle}{Proceedings of 14th Asian Congress of Fluid
  Mechanics (ACFM)}. volume~\bibinfo{volume}{2}; \bibinfo{year}{2013}. p.
  \bibinfo{pages}{515--521}.
\bibitem[{Ranjan et~al.(2014)Ranjan, Deshpande and
  Narasimha}]{ranjan2014direct}
\bibinfo{author}{Ranjan\xfnm[ R.]}, \bibinfo{author}{Deshpande\xfnm[ S.]},
  \bibinfo{author}{Narasimha\xfnm[ R.]}.
\newblock \bibinfo{title}{Direct numerical simulation of compressible flow past
  a low pressure turbine blade at high incidence}.
\newblock In: \bibinfo{booktitle}{ASME 2014 4th Joint US-European Fluids
  Engineering Division Summer Meeting collocated with the ASME 2014 12th
  International Conference on Nanochannels, Microchannels, and Minichannels}.
  \bibinfo{organization}{American Society of Mechanical Engineers};
  \bibinfo{year}{2014}. p. \bibinfo{pages}{V01AT02A010--V01AT02A010}.
\bibitem[{Ranjan et~al.(2015)Ranjan, Deshpande and Narasimha}]{rajesh2015IUTAM}
\bibinfo{author}{Ranjan\xfnm[ R.]}, \bibinfo{author}{Deshpande\xfnm[ S.M.]},
  \bibinfo{author}{Narasimha\xfnm[ R.]}.
\newblock \bibinfo{title}{A High-Resolution Compressible DNS Study of Flow Past
  a Low-Pressure Gas Turbine Blade}; \bibinfo{publisher}{World Scientific}.
\newblock p. \bibinfo{pages}{291--301}.
\newblock \URLprefix
  \url{http://www.worldscientific.com/doi/abs/10.1142/9789814635165_0028}.
  \DOIprefix\doi{10.1142/9789814635165_0028}.
\bibitem[{Rausch et~al.(1992)Rausch, BATINA and Yang}]{rausch1992spatial}
\bibinfo{author}{Rausch\xfnm[ R.D.]}, \bibinfo{author}{BATINA\xfnm[ J.T.]},
  \bibinfo{author}{Yang\xfnm[ H.T.]}.
\newblock \bibinfo{title}{Spatial adaptation of unstructured meshes for
  unsteady aerodynamic flow computations}.
\newblock \bibinfo{journal}{AIAA journal}
  \bibinfo{year}{1992};\bibinfo{volume}{30}(\bibinfo{number}{5}):\bibinfo{pages}{1243--1251}.
\bibitem[{Shoeybi et~al.(2010)Shoeybi, Sv{\"a}rd, Ham and Moin}]{shoeybi2010}
\bibinfo{author}{Shoeybi\xfnm[ M.]}, \bibinfo{author}{Sv{\"a}rd\xfnm[ M.]},
  \bibinfo{author}{Ham\xfnm[ F.E.]}, \bibinfo{author}{Moin\xfnm[ P.]}.
\newblock \bibinfo{title}{An adaptive implicit-explicit scheme for the dns and
  les of compressible flows on unstructured grids}.
\newblock \bibinfo{journal}{J Comput Phys}
  \bibinfo{year}{2010};\bibinfo{volume}{229}(\bibinfo{number}{17}):\bibinfo{pages}{5944--5965}.
\newblock \DOIprefix\doi{10.1016/j.jcp.2010.04.027}.
\bibitem[{Shu and Osher(1988)}]{shu1988efficient}
\bibinfo{author}{Shu\xfnm[ C.W.]}, \bibinfo{author}{Osher\xfnm[ S.]}.
\newblock \bibinfo{title}{Efficient implementation of essentially
  non-oscillatory shock-capturing schemes}.
\newblock \bibinfo{journal}{Journal of Computational Physics}
  \bibinfo{year}{1988};\bibinfo{volume}{77}(\bibinfo{number}{2}):\bibinfo{pages}{439--471}.
\bibitem[{Stadtmuller(2002)}]{stadtmuller1}
\bibinfo{author}{Stadtmuller\xfnm[ P.]}.
\newblock \bibinfo{title}{Investigation of wake-induced transition on the {LP}
  turbine cascade {T106A-EIZ}}.
\newblock \bibinfo{type}{Technical Report}; \bibinfo{year}{2002}.
\bibitem[{Subbareddy and Candler(2009)}]{subbareddy2009}
\bibinfo{author}{Subbareddy\xfnm[ P.K.]}, \bibinfo{author}{Candler\xfnm[
  G.V.]}.
\newblock \bibinfo{title}{A fully discrete, kinetic energy consistent
  finite-volume scheme for compressible flows}.
\newblock \bibinfo{journal}{Journal of Computational Physics}
  \bibinfo{year}{2009};\bibinfo{volume}{228}(\bibinfo{number}{5}):\bibinfo{pages}{1347--1364}.
\newblock \DOIprefix\doi{10.1016/j.jcp.2008.10.026}.
\bibitem[{Sutherland(1893)}]{sutherland1893lii}
\bibinfo{author}{Sutherland\xfnm[ W.]}.
\newblock \bibinfo{title}{Lii. the viscosity of gases and molecular force}.
\newblock \bibinfo{journal}{The London, Edinburgh, and Dublin Philosophical
  Magazine and Journal of Science}
  \bibinfo{year}{1893};\bibinfo{volume}{36}(\bibinfo{number}{223}):\bibinfo{pages}{507--531}.
\bibitem[{Van~Dyke(1962)}]{van1962higher}
\bibinfo{author}{Van~Dyke\xfnm[ M.]}.
\newblock \bibinfo{title}{Higher approximations in boundary layer theory}.
\newblock \bibinfo{journal}{J Fluid Mech}
  \bibinfo{year}{1962};\bibinfo{volume}{14}:\bibinfo{pages}{161--177}.
\bibitem[{Wei and Pollard(2011)}]{wei2011direct}
\bibinfo{author}{Wei\xfnm[ L.]}, \bibinfo{author}{Pollard\xfnm[ A.]}.
\newblock \bibinfo{title}{Direct numerical simulation of compressible turbulent
  channel flows using the discontinuous galerkin method}.
\newblock \bibinfo{journal}{Computers \& Fluids}
  \bibinfo{year}{2011};\bibinfo{volume}{47}(\bibinfo{number}{1}):\bibinfo{pages}{85--100}.
\bibitem[{Wissink(2003)}]{wissink2003}
\bibinfo{author}{Wissink\xfnm[ J.]}.
\newblock \bibinfo{title}{{DNS} of separating, low reynolds number flow in a
  turbine cascade with incoming wakes}.
\newblock \bibinfo{journal}{International Journal of Heat and Fluid Flow}
  \bibinfo{year}{2003};\bibinfo{volume}{24}(\bibinfo{number}{4}):\bibinfo{pages}{626--635}.
\newblock \DOIprefix\doi{10.1016/S0142-727X(03)00056-0}.
\bibitem[{Wissink and Rodi(2006)}]{wissink2006direct}
\bibinfo{author}{Wissink\xfnm[ J.]}, \bibinfo{author}{Rodi\xfnm[ W.]}.
\newblock \bibinfo{title}{Direct numerical simulations of transitional flow in
  turbomachinery}.
\newblock \bibinfo{journal}{Journal of turbomachinery}
  \bibinfo{year}{2006};\bibinfo{volume}{128}(\bibinfo{number}{4}):\bibinfo{pages}{668--678}.
\bibitem[{{Wissink} and {Rodi}(2006)}]{wissink_jfm2006}
\bibinfo{author}{{Wissink}\xfnm[ J.G.]}, \bibinfo{author}{{Rodi}\xfnm[ W.]}.
\newblock \bibinfo{title}{{Direct numerical simulation of flow and heat
  transfer in a turbine cascade with incoming wakes}}.
\newblock \bibinfo{journal}{Journal of Fluid Mechanics}
  \bibinfo{year}{2006};\bibinfo{volume}{569}:\bibinfo{pages}{209--247}.
\newblock \DOIprefix\doi{10.1017/S002211200600262X}.
\bibitem[{Wissink et~al.(2006)Wissink, Rodi and Hodson}]{wissink2006}
\bibinfo{author}{Wissink\xfnm[ J.G.]}, \bibinfo{author}{Rodi\xfnm[ W.]},
  \bibinfo{author}{Hodson\xfnm[ H.P.]}.
\newblock \bibinfo{title}{The influence of disturbances carried by periodically
  incoming wakes on the separating flow around a turbine blade}.
\newblock \bibinfo{journal}{International Journal of Heat and Fluid Flow}
  \bibinfo{year}{2006};\bibinfo{volume}{27}(\bibinfo{number}{4}):\bibinfo{pages}{721--729}.
\newblock \DOIprefix\doi{10.1016/j.ijheatfluidflow.2006.02.016}.
\bibitem[{Wu and Durbin(2001)}]{wu2001}
\bibinfo{author}{Wu\xfnm[ X.]}, \bibinfo{author}{Durbin\xfnm[ P.A.]}.
\newblock \bibinfo{title}{Evidence of longitudinal vortices evolved from
  distorted wakes in a turbine passage}.
\newblock \bibinfo{journal}{Journal of Fluid Mechanics}
  \bibinfo{year}{2001};\bibinfo{volume}{446}(\bibinfo{number}{1}):\bibinfo{pages}{199--228}.

\end{thebibliography}

\end{document}